\documentclass[12pt,english,longbibliography]{amsart}
\usepackage[utf8]{inputenc}
\usepackage[a4paper]{geometry}
\usepackage{fullpage}
\usepackage{ulem}
\usepackage{enumerate}
\usepackage{bm}
\usepackage{amsmath, amsthm, amsfonts, amssymb}

\def\thetitle{ Trend to Equilibrium for the Becker-D\"oring Equations: An Analogue of Cercignani's Conjecture }
\def\theauthor{Jos\'e A. Ca\~nizo, Amit Einav and Bertrand Lods}
\usepackage[colorlinks=true,linkcolor=blue,citecolor=blue,urlcolor=blue,breaklinks]{hyperref}
\hypersetup{pdfauthor={\theauthor}}

\usepackage{url}

\newcommand{\abs}[1]{\left\vert#1\right\vert}

\newcommand{\norm}[1]{\left\Vert#1\right\Vert}

\def\R{\mathbb{R}}

\def\N{\mathbb{N}}

\def\Q{\mathcal{Q}}

\def \leq {\leqslant}
\def \geq {\geqslant}

\def \d {\,\mathrm{d} }
\def \dt {\,\mathrm{d}t }

\def \dt {\,\mathrm{d}t }

\def \ddt {\frac{\d}{\dt}}

\def \vartheta {\varpi}

\def \zs {z_{\mathrm{s}}}

\DeclareMathOperator{\Ent}{\mathbf{Ent}}

\numberwithin{equation}{section}

\newcommand{\pa}[1]{\left(#1 \right)}
\newcommand{\avg}[1]{\langle #1 \rangle}
\newcommand{\br}[1]{\left\lbrace #1 \right\rbrace}

\newtheorem{thm}{Theorem}[section]
\newtheorem*{thm*}{Theorem}
\newtheorem{cor}[thm]{Corollary}
\newtheorem{lem}[thm]{Lemma}
\newtheorem{prp}[thm]{Proposition}
\newtheorem{hyp}{Hypothesis}

\theoremstyle{definition}
\newtheorem{dfn}[thm]{Definition}
\theoremstyle{remark}
\newtheorem{rem}[thm]{Remark}
\theoremstyle{example}

\title{\thetitle}

\date{}

\author{José A. Cañizo}
\address{José A. Cañizo, Departamento de Matem\'{a}tica
  Aplicada, Universidad de Granada, Av. Fuentenueva S/N, 18071
  Granada, Spain}
\email{canizo@ugr.es}
\thanks{JAC was supported by the Marie-Curie CIG grant KineticCF and
  the Spanish project MTM2014-52056-P}

\author{Amit Einav}
\address{Amit Einav, Department of Pure Mathematics and Mathematical Statistics,
  University of Cambrigde, United Kingdom}
\email{A.Einav@dpmms.cam.ac.uk}
\thanks{AE was supported by EPSRC grant EP/L002302/1.}

\author{Bertrand Lods}
\address{Bertrand Lods, Dipartement of Economics and Statistics  \& Collegio
  Carlo Alberto, Universit\`{a} degli Studi di Torino,  Corso Unione
  Sovietica, 218/bis, 10134 Torino, Italy}
\email{bertrand.lods@unito.it}

\begin{document}

\normalem

\maketitle
 
\begin{abstract}
  In this work we investigate the rate of convergence to equilibrium
  for subcritical solutions to the Becker-D\"oring equations with
  physically relevant coagulation and fragmentation coefficients and
  mild assumptions on the given initial data. Using a discrete version
  of the log-Sobolev inequality with weights we show that in the case
  where the coagulation coefficient grows linearly and the detailed
  balance coefficients are of typical form, one can obtain a linear
  functional inequality for the dissipation of the relative free
  energy. This results in showing Cercignani's conjecture for the
  Becker-D\"oring equations and consequently in an exponential rate of
  convergence to equilibrium. We also show that for all other typical
  cases one can obtain an 'almost' Cercignani's conjecture that
  results in an algebraic rate of convergence to
  equilibrium. Additionally, we show that if one assumes an
  exponential moment condition one can recover Jabin and Niethammer's
  rate of decay to equilibrium, i.e. an exponential to some fractional
  power of $t$.
\end{abstract}

\tableofcontents

\section{Introduction}
\label{sec:intro}

\subsection{The Becker-D\"oring Equations}
\label{sec: BD equations}

The Becker-D\"oring equations are a fundamental set of equations which
describe the kinetics of a first order phase transition. Amongst the
phenomena to which it applies one can find crystallisation, vapour
condensation, aggregation of lipids and phase separations in alloys
(see for instance \cite{Schmelzer05,Oxtoby}).

The Becker-D\"oring equations give the time evolution of the size
distribution of clusters of a certain substance. Denoting by
$\br{c_i(t)}_{i\in \N}$, the density of clusters of size $i$ at time
$t \geq 0$ (i.e. the density of clusters that are composed of $i$
particles), the equations read
\begin{subequations}
  \label{eq:BD}
  \begin{align}
    \label{eq:BD1}
    \frac{\d}{\d t} c_i(t) &= W_{i-1}(t) - W_{i}(t), \qquad i\in \N\setminus \br{1},
    \\
    \label{eq:BD2}
    \frac{\d}{\d t} c_1(t) &= - W_1(t) - \sum_{k=1}^\infty W_{k}(t),
  \end{align}
\end{subequations}
where
\begin{equation}
  \label{eq:def-Wi}
  W_{i}(t) := a_{i}\, c_1(t) c_i(t) - b_{i+1}\, c_{i+1}(t)
  \qquad i \in \N.
\end{equation}
and $a_i,b_i$, assumed to be strictly positive, are the
\emph{coagulation and fragmentation coefficients}. They determine the
rate at which clusters of size $i$ combine with clusters of size $1$
to create a cluster of size $i+1$, or clusters of size $i+1$ break
into clusters of size $i$ and $1$. This corresponds to the basic
assumption of the underlying model: if we represent symbolically by
$\{i\}$ the species of clusters of size $i$, then the only (relevant)
chemical reactions that take place are
\begin{equation*}
  \{i\} + \{1\} \rightleftharpoons \{i+1\}.
\end{equation*}

The quantity $W_i(t)$ defined in \eqref{eq:def-Wi} represents the
\emph{net rate} of the reaction
$\{i\} + \{1\} \rightleftharpoons \{i+1\}$, and under the above set of
equations it is easy to see that the \emph{density}, or \emph{mass},
of the solution, defined by
\begin{equation}
  \label{eq:def-mass}
  \varrho :=
  \sum_{i=1}^\infty i c_i(0)
  = \sum_{i=1}^\infty i c_i(t)
\end{equation}
is conserved under the Becker-D\"oring evolution. The original
equations proposed by Becker and D\"oring \cite{BD35} were similar to
\eqref{eq:BD}, with the slight change that the density of one particle
$c_1$, usually called the monomer density, was assumed to be
constant. The current version, motivated by the conservation of total
density, was first discussed in \cite{B77} and \cite{PL79} and is
widely used in classical nucleation theory. For more information about
the physical background and applications of the equations we refer the
interested reader to the aforementioned works as well as the recent
reviews \cite{Schmelzer05,Oxtoby}.

Much like in other kinetic equations, the study of a state of
equilibrium and the convergence to it is a fundamental question in
the study of the Becker-D\"oring equations. Defining the
\emph{detailed balance coefficients} $Q_i$ recursively by
\begin{equation}
  \label{eq:Qi}
  Q_1 = 1,
  \quad
  Q_{i+1} = \frac{a_i}{b_{i+1}} Q_i
  \quad i\in\N
\end{equation}
one can see that for a given $z\geq 0$ the sequence
\begin{equation}
  \label{eq:equilibria}
  c_i=Q_i z^i
\end{equation}
is formally an equilibrium of \eqref{eq:BD}. However, depending on the
coagulation and fragmentation coefficients $a_i$ and $b_i$, many of
these formal equilibria do not have a finite mass. The largest
$z_s \geq 0$, possibly $z_s=+\infty$, for which
\begin{equation*}
  \sum_{i=1}^{\infty} iQ_i z^i < +\infty
  \quad \text{ for all $0\leq z <z_s$}
\end{equation*}
is called the \emph{critical monomer density}, or sometimes the
monomer saturation density. The \emph{critical mass} (or, again,
saturation mass) is then defined by
\begin{equation}
  \label{eq:critical-mass}
  \varrho_s := \sum_{i=1}^\infty i Q_i z_s^i \in [0,+\infty].
\end{equation}

It is important to note that both $z_s$ and $\varrho_s$ are uniquely
determined by $a_i$ and $b_i$ and that $\br{Q_iz^i}_{i\in\N}$ is a
finite-mass equilibrium only for $0\leq z <z_s$, with the possibility
for the equality $z=z_s$ only when $\varrho_s < +\infty$.
Additionally, it is easy to see that for a given finite mass
$\varrho \leq \varrho_s$ there exists a unique $\overline{z}\geq 0$
such that
\begin{equation*}
\varrho=\sum_{i=1}^\infty iQ_i \overline{z}^i,
\end{equation*}
giving us a candidate for the asymptotic equilibrium state of
\eqref{eq:BD} under a given initial data. These are in fact the only
finite-mass equilibria (see \cite{BCP86}), and $\overline{z}$ defined
above is called the \emph{equilibrium monomer density} for a given
mass $\rho$.

A finite mass equilibrium is called \emph{subcritical} when its mass
$\varrho$, is strictly less than $\varrho_s$. It is called
\emph{critical} if $\varrho=\varrho_s$ and \emph{supercritical} if
$\varrho>\varrho_s$, assuming $\varrho_s<+\infty$. In this paper we
will only concern ourselves with subcritical solutions. Thus, to avoid
triviality we always assume that $z_s>0$.

The critical density $\varrho_s$, if finite, marks a change in the
behaviour of equilibrium states: if $\varrho<\varrho_s$ then a unique
equilibrium state with mass $\varrho$ exists, while if
$\varrho>\varrho_s$ no such equilibrium can occur and a phase
transition phenomenon takes place --- reflected in the fact that the
excess density $\varrho-\varrho_\mathrm{s}$ is concentrated in larger
and larger clusters as time progresses.

\subsection{Typical Coefficients}\label{sec:typical}

Physically motivated coagulation and fragmentation coefficients are
often given by
\begin{equation}
  \label{eq:PT-coefficients}
  a_i=i^\gamma,
  \qquad
  b_i = a_i \left(z_s+\dfrac{q}{i^{1-\mu}}\right),
  \qquad
  i\in\N,
\end{equation}
for some $0 < \gamma \leq 1$, $z_s >0$, $q >0$ and $0 <\mu < 1$ (see
\cite{Penrose89,N08} for details and concrete examples).

A different kind of reasoning, based on a statistical mechanics
argument involving the binding energy of clusters, results in the
coefficients
\begin{equation}
  \label{eq:CF-coefficients}
  a_i=i^\gamma,
  \qquad
  b_i = z_s (i-1)^\gamma
  \exp\big(\sigma i^\mu - \sigma (i-1)^\mu\big),
  \qquad
  i\in\N,
\end{equation}
for appropriate constants $\gamma,\mu$ and $\sigma$ (see for instance
\cite{JACBL,citeulike:4140416,citeulike:1069570,NCB02}). The behaviour
of \eqref{eq:PT-coefficients} and \eqref{eq:CF-coefficients} is
similar: for both of them we can write (by definition of $Q_i$)
 \begin{equation}\label{eq:Qi-form}
  Q_i
  = \dfrac{a_1 a_2 \ldots a_{i-1}}{b_2 b_3 \ldots b_i}
  = z_s^{1-i}\alpha_i,
\end{equation}
where $\br{\alpha_i}_{i\in \N}$ is non-increasing and satisfies 
 \begin{equation*}
 \lim_{i\rightarrow\infty}\frac{\alpha_{i+1}}{\alpha_{i}}=1.
\end{equation*}  
Our results are valid for both types of coefficients
\eqref{eq:CF-coefficients} and \eqref{eq:PT-coefficients}, which are
often used in the literature and cover a large range of applicable
cases.

\subsection{Previous Results}\label{sec: Known}

Let us briefly review existing results on the mathematical theory of
the Becker-D\"oring equations, which has advanced much since the first
rigorous works on the topic \cite{BCP86,BC88}. In \cite{BCP86} the
authors showed existence and uniqueness of a global solution to
\eqref{eq:BD} when
\begin{equation}
  \label{eq:hyp-existence}
  a_i \leq C_1 i,
  \quad
  b_i \leq  C_2 i,
  \qquad
  \sum_{i=1}^\infty i^2 c_i(0) < +\infty,
\end{equation}

for some constants $C_1, C_2 > 0$. As expected, under the above
assumptions the unique solution conserves mass (this is,
\eqref{eq:def-mass} holds rigorously). This basic existence theory is
applicable to all solutions we consider in this work.

The asymptotic behaviour of solutions to \eqref{eq:BD} is one of the
most interesting aspects of the equation. Supercritical behaviour,
while not dealt with in this work, has a particularly interesting link
to late-stage coarsening and has been studied extensively in
\cite{Pen97,Vel98,CGPV02,N08}. Asymptotic approximations of such
solutions have been developed in
\cite{citeulike:1069570,citeulike:2862238,citeulike:4140416}.

Regarding the subcritical regime, it was proved in \cite{BCP86,BC88}
that solutions with subcritical mass $\varrho$ approach the unique
equilibrium with this mass (determined by \eqref{eq:def-mass}). A
fundamental quantity in understanding this approach is the \emph{free
  energy}, $H( {\bm{c}})$, defined (at least formally) for any
sequence $\bm{c}=\br{c_i}_{i\in\N}$ by
\begin{equation}
  \label{eq:def-V}
  H(\bm{c}) := \sum_{i=1}^\infty c_i \left( \log \frac{c_i}{Q_i} - 1 \right).
\end{equation}

It can be shown that $H( {\bm{c}(t)})$ decreases along solutions
$\bm{c} = \bm{c}(t)$ to the Becker-D\"oring equations; in fact, for a
(strictly positive, suitably decaying for large $i$) solution
$\bm{c}(t) = \br{c_i(t)}_{i \in\N}$ of \eqref{eq:BD} we have
\begin{multline}
  \label{eq:H-thm-BD}
  \frac{\d}{\dt} H(\bm{c}(t))
  = - D(\bm{c}(t))
  \\
  := -\sum_{i=1}^\infty a_i Q_i
  \left(\frac{c_1 c_i}{Q_i} - \frac{c_{i+1}}{Q_{i+1}} \right)
  \left(\log \frac{c_1 c_i}{Q_i}
    - \log \frac{c_{i+1}}{Q_{i+1}} \right)
  \leq 0.
\end{multline}
This free energy is motivated by physical considerations and
constitutes a Lyapunov functional for our equation. Since it does not
have a definite sign we define a more natural candidate to measure the
distance of $\bm{c}(t)=\br{c_i(t)}_{i\in \N}$ to the
equilibrium. Using the notation
\begin{equation*}
\pa{\Q_z}_i=Q_i z^i
\end{equation*}
and denoting by $\Q=\Q_{\overline{z}}$, we can define the
\emph{relative} free energy as
\begin{equation}
  \label{eq:def-VFz}
  H(\bm{c}|\Q) := \sum_{i=1}^\infty c_i \left( \log \frac{c_i}{\overline{z}^i Q_i} - 1
  \right) + \sum_{i=1}^\infty \overline{z}^i Q_i
  = H(\bm{c}) - \log \overline{z} \sum_{i=1}^\infty i c_i
  + \sum_{i=1}^\infty \overline{z}^i Q_i.
\end{equation}
The relative free energy has the same time derivative as the free
energy, and thus the same monotonicity property
$$\dfrac{\d}{\d t}H(\bm{c}(t)|\Q)=-D(\bm{c}(t)) \qquad \forall t \geq 0,$$
where the \emph{free energy dissipation} $D$ is defined in
\eqref{eq:H-thm-BD}. The relative free energy also satisfies
\begin{itemize}
\item $H(\bm{c}|\Q) \geq 0 $, as can be seen by writing
  \begin{equation}
    \label{eq:Hpositive}
    H(\bm{c}|\Q) = \sum_{i=1}^\infty \Q_i \varphi\left(
      \frac{c_i}{\Q_i} \right),
    \quad \text{ with } \varphi(r) := r \log r - r + 1 \geq 0
  \end{equation}
\item $H(\bm{c}|\Q)=0$ if and only if
  $c_i = \Q_i = Q_i \overline{z}^i$ {for any $i \in \N$}, which is
  readily seen from \eqref{eq:Hpositive}.
\end{itemize}
This hints that $H(\bm{c}|\Q)$ is the right distance to
investigate. Indeed, while strictly speaking $H(\bm{c}|\Q)$ is not a
distance, it does control the $\ell^{1}$ distance between $\bm{c}$ and
$\Q$ by the celebrated Csisz\'ar-Kullback
inequality\footnote{Sometimes called Pinsker or Kullback-Pinsker
  inequality.}, which in our case translates to
\begin{equation}\label{eq:csiszar kullback ineq}
{\|\bm{c}-\Q\|_{\ell^{1}(\N)}=}\sum_{i=1}^\infty \abs{c_i-\Q_i} \leq \sqrt{2\varrho H(\bm{c}|\Q)}.
\end{equation}

The issue of estimating the rate of convergence to equilibrium of
subcritical solutions is the main concern of this paper. The first
result in this direction was the work \cite{JN03} by Jabin and
Niethammer, where they investigated the possibility of applying the
so-called \emph{entropy method} to the Becker-D\"oring equation. This
consists roughly in looking for functional inequalities between a
suitable Lyapunov functional of the equation (generally called the
entropy; it corresponds to the relative free energy in our case) and
its dissipation, so that one obtains a differential inequality that
estimates the rate of convergence to equilibrium. In the case of the
Becker-D\"oring equation, it was proved in \cite{JN03} that there
exists a constant $C>0$, depending only on the fixed parameters of the
problem and the initial data, such that
\begin{equation}\label{eq:JN inequality}
  D(\bm{c}) \geq C \frac{H(\bm{c}|\Q)}{\pa{\log H(\bm{c}|\Q)}^2},
\end{equation}
for all nonnegative sequences $\bm{c}$ with subcritical mass $\varrho$,
satisfying $\epsilon \leq c_1 \leq \zs - \epsilon$ for some
$\epsilon > 0$ and
\begin{equation*}
  \sum_{i=1}^\infty e^{\mu i} c_i = M^{\mathrm{exp}} < +\infty.
\end{equation*}
The constant $C$ depends on $\epsilon$ and $M^{\mathrm{exp}}$. This
result applies under resonable conditions on the coefficients $a_i$
and $b_i$; in particular, it applies to the coefficients
\eqref{eq:PT-coefficients} and \eqref{eq:CF-coefficients}. If we
consider now a solution $\bm{c} = \bm{c}(t)$ to \eqref{eq:BD}, we may
apply the inequality \eqref{eq:JN inequality} to $\bm{c}(t)$ as long
as $\bm{c}(t)$ satisfies the appropriate conditions, obtaining
\begin{equation*}
  \ddt H(\bm{c}(t)|\Q) = -D(\bm{c}(t))
  \leq
  - C \frac{H(\bm{c}(t)|\Q)}{\pa{\log H(\bm{c}(t)|\Q)}^2}.
\end{equation*}
Adding to this some additional considerations for the times $t$ for
which the inequality \eqref{eq:JN inequality} is not applicable to
$\bm{c}(t)$, one can deduce that $H(\bm{c}(t)|\Q$ is (essentially)
bounded above by the solution of the above differential inequality,
namely that
\begin{equation*}
  H(\bm{c}(t)|\Q) \leq H(\bm{c}(0)|\Q) e^{-K t^{\frac{1}{3}}}
\end{equation*}
for some $K > 0$. Using inequality \eqref{eq:csiszar kullback ineq},
this gives an almost-exponential rate of convergence to equilibrium
for subcritical solutions in the $\ell^1(\N)$ norm.

The question remained open of whether the convergence is in fact
exponential or not. Recently this has been answered positively by two
of the authors \cite{JACBL} through a different approach involving a
detailed study of the spectrum of the linearisation of equation
\eqref{eq:BD} around a subcritical equilibrium. This is an approach
with a strong analogy to results in the theory of the Boltzmann
equation; we refer to \cite{DMV11, JACBL} for more details on this
parallel. The idea of the argument is to use the inequality
\eqref{eq:JN inequality} when one is far from equilibrium. Then, once
we have reached a region which is close enough to equilibrium, the
linearised regime is dominant and one can use the spectral study of
the linearised operator in order to show that the convergence is in
fact exponential. The outcome of this strategy is the following: for
many interesting coefficients (including \eqref{eq:PT-coefficients}
and \eqref{eq:CF-coefficients}), subcritical solutions $\bm{c} =
\bm{c}(t)$ to \eqref{eq:BD} with
\begin{equation*}
  \sum_{i=1}^\infty e^{\mu i} c_i(0) := M^{\mathrm{exp}} < +\infty
  \quad \text{for some $\mu > 0$}
\end{equation*}
satisfy that
$$\sum_{i=1}^\infty e^{{\mu'}i} \abs{c_i(t)-\Q_i} \leq
Ce^{-\lambda t}
\quad \text{for $t \geq 0$}
$$
for some $0 < \mu' < \mu$, $C > 0$ and $\lambda > 0$ which depend on
the parameters of the problem and on $M^{\mathrm{exp}}$. In fact,
$\mu$ and $C$ only depend on the initial data $\bm{c}(0)$ through its
mass and the value of $M^{\mathrm{exp}}$; $\lambda$ depends only on
the coefficients and the initial mass and can be estimated
explicitly. The value of $\lambda$ is bounded above by (and can be
taken very close to) the size of the spectral gap of the linearised
operator. Recently Murray and Pego \cite{MurrayPego} have used this
spectral gap and developed the local estimates of the linearised
operator in order to obtain convergence to equilibrium at a polynomial
rate with milder conditions on the decay of the initial data. These
results, like those in \cite{JACBL}, are local in nature and require
the use of some global estimate such as \eqref{eq:JN inequality} in
order to provide global rates of convergence to equilibrium.

\subsection{Main Results}\label{sec: main results}

Our main goal in this work is to complete the picture of convergence
to equilibrium by investigating modified and improved versions of the
inequality \eqref{eq:JN inequality}. We show optimal inequalities and
settle the question of whether full exponential convergence can be
obtained through a \emph{linear} inequality of the form
\begin{equation*}
  D(\bm{c}) \geq K H(\bm{c}|\Q).
\end{equation*}
in some cases. In analogy to the Boltzmann equation, we refer to the
question of whether such $K$ exists along solutions to \eqref{eq:BD}
as \emph{Cercignani's conjecture for the Becker-D\"oring equations.}
In fact, we show that under relatively mild conditions on the initial
data, typical coagulation and fragmentation coefficients (such as
\eqref{eq:PT-coefficients} and \eqref{eq:CF-coefficients}) admit an
``almost'' Cercignani conjecture for the energy dissipation, i.e. an
inequality bounding below $D(\bm{c})$ by a power of $H(\bm{c}|\Q)$,
yielding an explicit rate of convergence to equilibrium. Surprisingly,
we also find a relevant case ($a_i \sim i$ for all $i$) for which the
conjecture is actually valid.

We will often require the following assumptions on the coagulation and
fragmentation coefficients. Some of these are similar to those in
\cite{JN03}, and always include coefficients of the form
\eqref{eq:PT-coefficients} and \eqref{eq:CF-coefficients}. We recall
that we always assume $a_i, b_i > 0$ for all $i \in \N$, and that the
detailed balance coefficients $Q_i$ were defined in \eqref{eq:Qi} ---
given $a_i$ one can determine $b_i$ through $Q_i$, and vice versa.
\begin{hyp}\label{hyp: z_s>0}
  $0<z_s<+\infty$.
\end{hyp}

\begin{hyp}\label{hyp: Qi-form}
  For all $i\in \N$, $Q_i=z_s^{1-i}\alpha_i$, where
  $\br{\alpha_i}_{i\in\N}$ is a non-increasing positive sequence,
  $\alpha_1>0$ and
  $\lim_{i\rightarrow\infty}\frac{\alpha_{i+1}}{\alpha_{i}}=1$.
\end{hyp}

\begin{hyp}\label{hyp: condition on a_i}
  $a_i=O(i^\gamma)$ for some $0<\gamma\leq 1$, i.e. there exist
  $C_1,C_2>0$ such that
  \begin{equation*}
    C_1 i^\gamma \leq a_i \leq C_2 i^\gamma
    \quad
    \text{ for all $i\in\N$.}
  \end{equation*}
\end{hyp}

In most of the estimates we obtain, a crucial role will be played by
the \emph{lower free energy dissipation}, $\overline{D}(\bm{c})$,
defined for a given non-negative sequence $\bm{c}$ by 
\begin{equation}\label{eq:lower free energy}
  \overline{D}(\bm{c})
  =
  \sum_{i=1}^\infty a_i Q_i\pa{\sqrt{\frac{c_1c_i}{Q_i}}
    - \sqrt{\frac{c_{i+1}}{Q_{i+1}}}}^2
\end{equation}
At this point one notices that the elementary inequality
$(x-y)(\log x-\log y) \geq \left(\sqrt{x}-\sqrt{y}\right)^{2}$ when
$x,y>0$ implies that
$$D(\bm{c}) \geq \overline{D}(\bm{c})$$
for any non-negative sequence $\bm{c}$. Thus, any lower bound that is
obtained for $\overline{D}(\bm{c})$ will transfer immediately to
$D(\bm{c})$.

\medskip
We now state our main result on general functional inequalities for
the free energy dissipation, from which later we conclude a
quantitative rate of convergence to equilibrium. It can be divided in
two parts: functional inequalities when $c_1$ is not too small, nor is
too far from $\overline{z}$, and inequalities in the case where $c_1$
escapes the above region.

\begin{thm}\label{thm:main thm fe}
  Let $\br{a_i}_{i\in\N},\br{Q_i}_{i\in\N}$ satisfy Hypotheses
  \ref{hyp: z_s>0}-\ref{hyp: condition on a_i} with
  $0\leq \gamma\leq 1$ and let $\bm{c}=\br{c_i}_{i\in\N}$ be an
  arbitrary positive sequence with finite total density
  $0 < \varrho < \varrho_{\mathrm{s}}$.
  \begin{enumerate}[(i)]
  \item \label{enm: gamma=1 cerc conjecture}

    \textbf{(Estimate for $a_i \sim i$.)}
    Assume that $\gamma=1$ and that there exist $\delta> 0$ such that
    \begin{equation}\label{eq:good c_1}
      \delta < c_1 < z_s - \delta.
    \end{equation} 
    Then there exists $K>0$ depending only on $\delta, \zs, \varrho$
    and $\br{\alpha_i}_{i\in\N}$ such that
    \begin{equation}\label{eq:cerc conjecture BD}
      \overline{D}(\bm{c}) \geq K H(\bm{c}|\Q).
    \end{equation}
    
    \medskip
  \item \label{enm: gamma<1 almost cerc conjecture}

    \textbf{(Estimate for $a_i \sim i^\gamma$ with $\gamma < 1$.)}
    Assume that $0 \leq \gamma < 1$ and that $c_1$ satisfies
    \eqref{eq:good c_1} for some $\delta > 0$. If, in
    addition, there exists $\beta>1$ with
    \begin{equation}\label{eq:Mbeta}
      M_\beta(\bm{c})=\sum_{i=1}^\infty i^\beta c_i < +\infty
    \end{equation}
    then there exists $K>0$ depending only on
    $\delta, \zs, \varrho, M_\beta(\bm{c})$ and
    $\br{\alpha_i}_{i\in\N}$ such that
    \begin{equation}\label{eq:almost cerc conjecture fe}
      \overline{D}(\bm{c}) \geq K H(c|\Q)^{\frac{\beta-\gamma}{\beta-1}}.
    \end{equation}
    
  \item \label{enm: uniform lower bound}

    \textbf{(Estimate for small $c_1$.)}  Assume that $\gamma = 1$, or
    that $0 \leq \gamma < 1$ and \eqref{eq:Mbeta} holds for some
    $\beta > 1$. Assume also that for some $\delta > 0$
    \begin{equation*}
      c_1 \leq \delta
    \end{equation*}
    or that
    \begin{equation*}
      c_1 \geq z_{\mathrm{s}} - \delta
    \end{equation*}
    (i.e., $c_1$ is outside of the range given in \eqref{eq:good
      c_1}). Then if $\delta > 0$ is small enough (depending only on
    $\varrho$ and $\{Q_i\}_{i \in \N}$), there exists $\varepsilon>0$
    depending only on $\delta, \zs, \varrho$ and
    $\br{\alpha_i}_{i\in\N}$ if $\gamma=1$ (and additionally on
    $M_\beta(\bm{c})$ if $\gamma<1$) such that
    \begin{equation}\label{eq: uniform lower bound on energy dissipation}
      \overline{D}(\bm{c}) \geq \varepsilon.
    \end{equation}
  \end{enumerate}
  The constants $K$ and $\varepsilon$ can be estimated explicitly in
  all cases.
  
\end{thm}

We point out that since $\overline{z}$, $\zs$, $\varrho_{\mathrm{s}}$,
$\br{Q_i}_{i\in\N}$ and $\{\alpha_i\}_{i \in \N}$ are determined
entirely by the coagulation and fragmentation coefficients and
$\varrho$, all constants in the above theorem depend only on
$\varrho$, the coefficients $\{a_i\}_{i\in\N}$, $\{b_i\}_{i\in\N}$,
and the additional bounds $\delta$ or $M_\beta$.

The case \eqref{enm: gamma=1 cerc conjecture} of Theorem \ref{thm:main
  thm fe} is optimal in the following sense:
\begin{thm}
  \label{thm:optimal}
  Call $X_{\varrho}$ the set of
  nonnegative sequences $\bm{c} = \{c_i\}_{i \in \N}$ with mass
  $\varrho$ (i.e., such that $\sum_{i=1}^\infty i c_i =
  \varrho$). Then, there exist $\br{a_i}_{i\in\N},\br{Q_i}_{i\in\N}$ that satisfy Hypotheses
  \ref{hyp: z_s>0}-\ref{hyp: condition on a_i} with $\gamma < 1$ such that
  \begin{equation*}
    \inf_{X_\varrho} \frac{D(\bm{c})}{H(\bm{c}|\Q)} = 0.
  \end{equation*}
  for any $\varrho < \varrho_{\mathrm{s}}$.
\end{thm}
Of course, this shows that a linear inequality as that of
Theorem \ref{thm:main thm fe} (\ref{enm: gamma=1 cerc conjecture}) cannot hold
if $a_i \sim i^\gamma$ with $\gamma < 1$.

\medskip The idea behind the proof of Theorem \ref{thm:main thm fe} is
to use a discrete logarithmic Sobolev inequality with weights,
motivated by works of Bobkov and G\"otze \cite{BG99} and Barthe and
Roberto \cite{Barthe}, to show part \eqref{enm: gamma=1 cerc
  conjecture}. As the conditions for the validity of the log-Sobolev
inequality are \emph{not} satisfied under the conditions of part
\eqref{enm: gamma<1 almost cerc conjecture}, a simple interpolation is
used to show the desired result in that case. Part \eqref{enm: uniform
  lower bound} is proved by two estimates: The case where $c_1$ is too large follows an idea of Jabin and Niethammer, and is essentially stated
already in \cite{JN03}, while the case where $c_1$ is too small seems to be a new result which we provide.

\medskip
From Theorem \ref{thm:main thm fe} one can conclude in a
straightforward way the following theorem, our main result on the rate
of convergence to equilibrium:

\begin{thm}\label{thm:rate of convergence}
  Let $\br{a_i}_{i\in\N},\br{Q_i}_{i\in\N}$ satisfy Hypotheses
  \ref{hyp: z_s>0}--\ref{hyp: condition on a_i} with
  $0\leq \gamma \leq 1$, and let $\bm{c}(t)=\br{c_i(t)}_{i\in\N}$ be a
  solution to the Becker-D\"oring equations with mass
  $\varrho \in (0,\varrho_{s})$.
  \begin{enumerate}[(i)]
  \item \label{enm:rate of convergence gamma=1} \textbf{(Rate for $a_i \sim i$.)} If $\gamma=1$ then there
    exists a constant $K>0$ depending only on $\zs,\varrho $ and
    $\br{\alpha_i}_{i\in\N}$, and a constant $C > 0$
    depending only on $H(\bm{c}(0)|\Q),\zs,\varrho$ and
    $\br{\alpha_i}_{i\in\N}$ such that
    \begin{equation*}
      H(\bm{c}(t)|\Q) \leq C e^{-Kt}
      \quad \text{ for $t \geq 0$.}
    \end{equation*}
  \item \label{enm:rate of convergence gamma<1} \textbf{(Rate for $a_i \sim i^\gamma$, $\gamma < 1$.)} If
    $\gamma<1$ and $M_\beta(\bm{c}(0))<+\infty$ then there exists a
    constant $K>0$ depending only on $\zs,\varrho, M_\beta$
    and $\br{\alpha_i}_{i\in\N}$, and a constant
    $C > 0$ depending only on
    $H(\bm{c}(0)|\Q), \zs, \varrho, M_\beta$ and
    $\br{\alpha_i}_{i\in\N}$ such that
    \begin{equation*}
      H(\bm{c}(t)|\Q)
      \leq
      \frac{1}{\pa{C
          +\frac{1-\alpha}{\beta-1}Kt}^{\frac{\beta-1}{1-\alpha}}}
      \quad \text{ for $t \geq 0$.}
    \end{equation*}
  \end{enumerate}
  The constants $K$ and $C$ can be estimated explicitly.
\end{thm}
As remarked above, the constants $C$ and $K$ above depend ultimately
only on the coefficients $a_i$, $b_i$, the initial mass $\varrho$, and
the moment $M_\beta$ in the case \emph{(ii)}.

There are some improvements in these theorems with respect to the
existing theory. One of them is that they apply to more general
initial conditions, removing the need for a finite exponential moment
present in \cite{JN03,JACBL}. Another one is that they answer the
question of whether one can obtain a linear inequality such as
\eqref{eq:cerc conjecture BD} (i.e., whether the equivalent of
Cercignani's conjecture holds), making clear the link to discrete
logarithmic Sobolev inequalities. Surprisingly, it does hold in the
case $a_i \sim i$, which is physically relevant for example in
modelling polymer chains \cite{NCB02,citeulike:4140416}. As a result,
the statement for $a_i \sim i$ is quite strong: it gives full
exponential convergence, with explicit constants in terms of the
parameters, with no restriction on the initial data except that of
subcritical mass. Point \eqref{enm:rate of convergence gamma<1} in \ref{thm:rate of convergence}
also relaxes the requirements on the initial data, at the price of
obtaining a slower convergence than that of \cite{JACBL}; we do not
know whether this rate is optimal for initial conditions with
polynomially decaying tails (so that $M_\beta < \infty$ for some
$\beta > 1$, but $M_{\beta'} = +\infty$ for some $\beta' > \beta$).

One may wonder if the method presented here can be used to reach an
inequality like Jabin and Neithammer's \eqref{eq:JN inequality} under
the additional condition of an exponential moment. The answer is
provided in the following:
\begin{thm}\label{thm:rate of convergence exp}
  Let $\br{a_i}_{i\in\N},\br{Q_i}_{i\in\N}$ satisfy Hypothesis
  \ref{hyp: z_s>0}--\ref{hyp: condition on a_i} with $0\leq \gamma
  <1$.
  Let $\bm{c}=\br{c_i}_{i\in\N}$ be an arbitrary positive sequence
  with mass $\varrho\in(0,\varrho_{s})$ for which there exists $\mu>0$
  such that
\begin{equation}\label{eq:almost cerc exp condition fe}
  {M^{\exp}_\mu(\bm{c})}=\sum_{i=1}^\infty e^{\mu i} c_i < +\infty.
\end{equation}
Then:
\begin{enumerate}[(i)]
 \item \label{enm:exp func ineq} \textbf{(functional inequality.)} There exist $K_1,K_2,\varepsilon>0$ depending only on
$\zs,\varrho, M^{\exp}_\mu(\bm{c})$ and $\br{\alpha_i}_{i\in\N}$ such
that 
\begin{equation}\label{eq:almost cerc conjecture exp fe}
  \overline{D}(\bm{c}) \geq
  \min\pa{\frac{K_1H(\bm{c}|\Q)}{\abs{\log\pa{K_2
          H(\bm{c}|\Q)}}^{1-\gamma}},\varepsilon}.
\end{equation}
Moreover, $K_1,K_2$ and $\varepsilon$ can be given explicitly.
\item \label{enm:exp mom rate} \textbf{(rate of convergence.)}
If $\bm{c}(t)=\br{c_i(t)}_{i\in\N}$ is a solution to
the Becker-D\"oring equations with mass $0<\varrho<\varrho_{s}$ such that there exists $\mu>0$ with
$ {M^{\exp}_\mu(\bm{c}(0))}<+\infty,$ then there exists a constant $K>0$ depending only on
$\zs,\varrho, {M}^{\exp}_{\mu}(\bm{c}(0))$ and
$\br{\alpha_i}_{i\in\N}$, and a constant $C > 0$
depending only on
$H(\bm{c}(0)|\Q),\zs,\varrho, {M}^{\exp}_{\mu}(\bm{c}(0))$ and
$\br{\alpha_i}_{i\in\N}$ such that
\begin{equation*}
  H(\bm{c}(t)|\Q) \leq C  e^{-Kt^{\frac{1}{2-\gamma}}}.
\end{equation*}
Moreover, $K$ and $C$ can be given explicitly.
\end{enumerate}
\end{thm}

\subsection{Organisation of the Paper}
\label{sec: organisation}

The structure of the paper is as follows: In Section \ref{sec:discrete
  log-sob} we will present our main technical tool, a discrete version
of the log-Sobolev inequality with weights. {Section \ref{sec:EDI} contains the proof of Theorem \ref{thm:main thm fe}
  and uses Section \ref{sec:discrete log-sob} to show the first part
  of the theorem. We also show in this section that this method is
  optimal and that Cercignani's conjecture cannot hold when
  $\gamma<1$, proving Theorem \ref{thm:optimal} and explore the additional inequality that appears under the assumption of a finite exponential moment. Section \ref{sec:
    energy dissipation for BD} deals with the consequences of our
  functional inequalities for the solutions to the Becker-D\"oring
  equation and contains the proof of Theorem \ref{thm:rate of
    convergence} and part \eqref{enm:exp mom rate} of Theorem \ref{thm:rate of convergence exp}. In Section \ref{sec:application to general CF}
  we briefly point out some consequences of our results for general
  coagulation and fragmentation equations and remark on the
  difficulties of obtaining stronger results in this general
  setting. Lastly, we conclude this work with Section \ref{sec:final
    remarks} where we discuss a few final remarks, followed by
  appendices where we give proofs to some technical lemmas.}

\section{A Discrete Weighted Logarithmic Sobolev Inequality}
\label{sec:discrete log-sob}

The key ingredient in proving Cercignani's conjecture for the Becker-D\"oring equations in the setting of Theorem \ref{thm:main thm fe} relies heavily on a discrete log-Sobolev inequality with weights. The theory presented here follows closely the work done by Bobkov and G\"otze in \cite{BG99}, and that of Barthe and Roberto in \cite{Barthe}, and can be seen as a simple discrete version of the aforementioned papers. We will give a sketch of the proof here, and in Appendix \ref{app:discrete log-sob}, for the sake of completion. The reader that is familiar with these works is advised to skip this subsection and see how we use the main result proved in it in the next subsection.
\begin{dfn}\label{dfn:entropy}
We say that $\bm{\mu}\in P\pa{\N}$ if $\bm{\mu}=\br{\mu_i}_{i\in\N}$ is a non-negative sequence such that
\begin{equation*}
\sum_{i=1}^\infty \mu_i = 1.
\end{equation*}
For any non-negative sequence $\bm{g} = \br{g_i}_{i \in\N}$ with 
$$\sum_{i=1}^\infty \mu_i g_i < + \infty$$
 we define its \emph{entropy} with respect to $\bm{\mu}$ as
\begin{equation}
  \label{eq:def-entropy}
  \Ent_{\mu}(\bm{g}) = \sum_{i=1}^\infty \mu_i g_i \log \frac{g_i}{
    \sum_{i=1}^\infty \mu_i g_i}.
\end{equation}
\end{dfn}

\begin{dfn}\label{dfn: log-sob}
Given $\bm{\mu}\in P\pa{\N}$ and positive sequence  $\bm{\nu} = \br{\nu_i}_{i\in \N}$ (not necessarily normalised) we say that $\bm{\nu}$ admits a log-Sobolev inequality with respect to $\mu$ with constant $0<\Lambda<+\infty$ if for any sequence $\bm{f}=\br{f_i}_{i\in\N}$
\begin{equation}\label{eq: log-sob}
\Ent_{\mu}\pa{\bm{f}^2} \leq \Lambda \sum_{i=1}^\infty \nu_i \pa{f_{i+1}-f_i}^2, 
\end{equation}
where $\bm{f}^2=\br{f_i^2}_{i\in\N}$.
\end{dfn}
In what follows we will always assume that $\bm{\mu}\in P\pa{\N}$.
Denoting by 
$$\Psi(x)=\abs{x}\log\pa{1+\abs{x}}$$
the main theorem, and its simplified corollary, that we will prove in this subsection are:
\begin{thm}
  \label{thm: log sobolev condition}
The following two conditions are equivalent:
  \begin{enumerate}[(i)]
  \item $\bm{\nu}$ admits a log-Sobolev inequality with respect to $\bm{\mu}$ with constant $\Lambda$.
  \item For any $m\in\N$ such that
    \begin{equation*}
    \max\pa{\sum_{i=1}^{m-1} \mu_i , \sum_{i=m+1}^\infty \mu_i}<\frac{2}{3}
    \end{equation*}
    we have that
    \begin{equation}\label{eq: log sobolev condition II}
   {B_1=\sup_{k\geq m}\dfrac{\sum_{i=1}^k \frac{1}{\nu_i}}{\Psi^{-1}\pa{\dfrac{1}{\sum_{i=k+1}^\infty \mu_i}}}<+\infty.}
   \end{equation}
  \end{enumerate}
  Moreover, if $(ii)$ is valid then one can choose 
  \begin{equation}\label{eq: choice of lambda}
   \Lambda=40(B_2+4B_1),
   \end{equation}
   where   
   ${B_2=\dfrac{\sum_{i=1}^{m-1}\frac{1}{\nu_i}}{\Psi^{-1}\pa{\frac{1}{\sum_{i=1}^{m-1} \mu_i}}}}.$
\end{thm}
\begin{cor}\label{cor: log sobolev equivalent}
The following two conditions are equivalent:
  \begin{enumerate}[(i)]
  \item  $\bm{\nu}$ admits a log-Sobolev inequality with respect to $\bm{\mu}$ with constant $\Lambda$.
  \item For any $m\in\N$ such that
    $$\max\pa{\sum_{i=1}^{m-1} \mu_i , \sum_{i=m+1}^\infty \mu_i}<\frac{2}{3}$$
    we have that
    \begin{equation}\label{eq: simple log sobolev condition II}
      D_1=\sup_{k\geq m}\pa{ -\sum_{i=k+1}^\infty \mu_i \log \pa{\sum_{i=k+1}^\infty \mu_i}}\pa{\sum_{i=1}^k \frac{1}{\nu_i}}<\infty.
    \end{equation}
  \end{enumerate}
Moreover, if $(ii)$ is valid then one can choose
   \begin{equation}\label{eq: simple choice of lambda}
   \Lambda=120(D_2+4D_1),
   \end{equation}
   where $D_2=\pa{ -\sum_{i=1}^{m-1} \mu_i \log \pa{\sum_{i=1}^{m-1} \mu_i}}\pa{\sum_{i=1}^{m-1}\frac{1}{\nu_i}}$.
\end{cor}
\begin{rem}
One can clearly see that if 
$$\sup_{k\geq 1}\pa{ -\sum_{i=k+1}^\infty \mu_i \log \pa{\sum_{i=k+1}^\infty \mu_i}}\pa{\sum_{i=1}^k \frac{1}{\nu_i}}<\infty$$
then one has a log-Sobolev inequality of $\bm{\nu}$ with respect to $\bm{\mu}$. However, the introduction of the 'approximate median' $m$ allows us to have an explicit estimation on the log-Sobolev constant $\Lambda$.
\end{rem}
The rest of the subsection is dedicated to the proof of the above theorem and corollary.
\begin{dfn}\label{dfn: L(c)}
Let $\bm{\mu} \in P\pa{\N}$. Given a sequence $\bm{f}=\br{f_i}_{i\in \N}$ we define
\begin{equation}\label{eq:L(c)}
\mathcal{L}(\bm{f})=\sup_{\alpha\in \R}\Ent_{\mu}\pa{(\bm{f}+\alpha )^2}
\end{equation}
where $\bm{f}+\alpha =\br{f_i+\alpha}_{i\in \N}$.
\end{dfn} 
\begin{lem}\label{lem:comparison between L and Entropy}
For any sequence $\bm{f}$, we have
\begin{equation}\label{eq:comparision between L and Entropy}
\Ent_{\mu}(\bm{f}^2)\leq \mathcal{L}(\bm{f}) \leq \Ent_{\mu}(\bm{f}^2)+2\sum_{i=1}^\infty \mu_i f_i^2. 
\end{equation}
\end{lem}
\begin{rem} 
This Lemma is an adaptation of the appropriate Lemma in \cite{Roth85}. We leave the proof of it to Appendix \ref{app:discrete log-sob}.
\end{rem}
The next step in our path is to recast the log-Sobolev inequality as a Poincar\'e inequality in the Orlicz space associated to $\Psi$.
\begin{dfn}\label{dfn:orlicz}
Given $\bm{\mu}\in P\pa{\N}$ and a Young Function, $\Sigma:[0,+\infty)\rightarrow [0,+\infty)$, i.e. a convex function such that
$$\frac{\Sigma(x)}{x}\underset{x \rightarrow +\infty}{\longrightarrow}+\infty, \quad \frac{\Sigma(x)}{x}\underset{x \rightarrow 0}{\longrightarrow}0, $$
we define the Orlicz space $L^{\pa{\mu}}_\Sigma$ as the space of all sequences $\bm{f}$ such that there exists $k>0$ with 
$$\sum_{i=1}^\infty \mu_i \Sigma\pa{\frac{\abs{f_i}}{k}} < \infty.$$
In that case we define
$$\norm{\bm{f}}_{L^{\pa{\mu}}_\Sigma}= \inf_{k>0} \br{\sum_{i=1}^\infty \mu_i \Sigma\pa{\frac{\abs{f_i}}{k}} \leq 1}.$$
\end{dfn}
In what follows we will drop the superscript $\mu$ from the Orlitz space of $\Psi$ and its norm. Additionally we denote by $\Phi(x)=\Psi(x^2)$ and notice that:
\begin{equation}\label{eq: observation about L_Phi and L_Psi}
\norm{\bm{f}^2}_{L_\Psi}=\inf_{k>0}\br{\sum_{i=1}^\infty \mu_i \Psi\pa{\frac{f^2}{k}} \leq 1} = \pa{\inf_{\sqrt{k}>0}\br{\sum_{i=1}^\infty \mu_i \Phi\pa{\frac{\abs{f}}{\sqrt{k}}} \leq 1}}^2=\norm{\bm{f}}_{L_\Phi}^2. 
\end{equation}
\begin{thm}\label{thm:log sobelev and poincare I}
The following conditions are equivalent:
  \begin{enumerate}[(i)]
  \item  $\bm{\nu}$ admits a log-Sobolev inequality with respect to $\bm{\mu}$ with constant $\Lambda$.
  \item For any sequence $\bm{f}$
  \begin{equation}\label{eq:log sobolev and poincare I with L}
  \mathcal{L}(\bm{f}) \leq \Lambda \sum_{i=1}^\infty \nu_i \pa{f_{i+1}-f_i}^2.
  \end{equation}
  \item For any sequence $\bm{f}$
    \begin{equation}\label{eq:log sobolev and poincare I with L_Phi}
      \norm{\bm{f}-\avg{\bm{f}}}_{L_\Phi}^2 \leq \lambda \sum_{i=1}^\infty \nu_i \pa{f_{i+1}-f_i}^2.
    \end{equation}
    where  $\avg{\bm{f}}=\sum_{i=1}^\infty \mu_i f_i$.
  \end{enumerate}
  Moreover, if $(i)$ or $(ii)$ are valid one can choose $\lambda=\frac{3}{2}\Lambda$. If $(iii)$ is valid one can choose $\Lambda=5\lambda$.
\end{thm}
The proof of the theorem relies on the following proposition:
\begin{prp}\label{prop:poincare and L}
For any sequence $\bm{f}$ one has that
\begin{equation}\label{eq:poincare and L}
\frac{2}{3}\norm{\bm{f}-\left\langle \bm{f} \right\rangle}^2_{L_\Phi} \leq \mathcal{L}(\bm{f}) \leq 5 \norm{\bm{f}-\left\langle \bm{f} \right\rangle}^2_{L_\Phi} 
\end{equation}
\end{prp}
\begin{proof}
We start by noticing that we may assume that $\avg{\bm{f}}=0$ as well as $\norm{\bm{f}-\avg{\bm{f}}}_{L_\Phi}=1$. This is true as $\mathcal{L}$ is invariant under translations and
$$\Ent_\mu(\alpha \bm{f}) = \alpha \Ent_{\mu}(\bm{f}).$$
Using Lemma \ref{lem:comparison between L and Entropy}, we find that
\begin{equation*}\begin{split}
\mathcal{L}(\bm{f}) &\leq \Ent_{\mu}(\bm{f}^2)+2\sum_{i=1}^\infty \mu_i f_i^2=\sum_{i=1}^\infty \mu_i f_i^2 \log \pa{f_i^2} + 2\sum_{i=1}^\infty \mu_i f_i^2 \\
&\phantom{+++++} - \pa{\sum_{i=1}^\infty \mu_i f_i^2}\log\pa{\sum_{i=1}^\infty \mu_i f_i^2}\\
&\leq \sum_{i=1}^\infty \mu_i  \Phi(f_i) +h\pa{\sum_{i=1}^\infty \mu_i f_i^2},\end{split}\end{equation*}
where $h(x)=2x-x\log x$ for $x > 0$. As $h$ is an increasing function on $[0,e]$ and 
\begin{equation*}
\norm{\bm{f}}_{L^1_\mu}\leq \norm{\bm{f}}_{L^2_\mu} \leq \sqrt{\frac{3}{2}} \norm{\bm{f}}_{L_\Phi},
\end{equation*}
(see Lemma \ref{lemapp: 1 2 and phi} in Appendix \ref{app:discrete log-sob}) we have that 
$$\norm{\bm{f}}^2_{L^2_\mu} \leq 2.$$ 
Thus, as
$$\sum_{i=1}^\infty \mu_i \Phi(f_i)= \sum_{i=1}^\infty \mu_i \Phi\pa{\frac{f_i}{\norm{f}_{L_\Phi}}} \leq 1,$$
we find that
$$\mathcal{L}(\bm{f}) \leq 1+h(2) \leq 5,$$
proving the right hand side inequality of (\ref{eq:poincare and L}). To show the left hand side inequality we assume that $\mathcal{L}(\bm{f})=2$. By the definition of $\mathcal{L}$ and the fact that
\begin{equation*}
\norm{\bm{f}-\avg{\bm{f}}}_{L^2_\mu}^2 = \frac{1}{2}\lim_{\abs{a}\rightarrow\infty}\Ent_\mu\pa{(\bm{f}+a)^2}
\end{equation*}
(see Lemma \ref{lemapp: entropy and L_2} in Appendix \ref{app:discrete log-sob}) we know that
$$\norm{\bm{f}}_{L^2_\mu}^2 \leq \frac{1}{2}\mathcal{L}(\bm{f})=1.$$
This implies that 
\begin{equation*}\begin{split}\sum_{i=1}^\infty \mu_i \Phi( f_i) &\leq 1+ \sum_{i=1}^\infty \mu_i f_i^2 \log f_i^2 = 1+ \Ent_{\mu}(\bm{f}^2)+\norm{\bm{f}}_{L^2_\mu}^2\log\pa{\norm{\bm{f}}_{L^2_\mu}^2}\\
&\leq 1+\mathcal{L}(\bm{f})=3,\end{split}\end{equation*}
where we have used the fact that $x\log\pa{1+x} \leq 1+x\log x$ when $x>0$. 

Since for any $a\geq 1$, $\Phi\pa{\frac{x}{\sqrt{a}}} = \frac{x^2}{a^2}\log\pa{1+\frac{x^2}{a^2}} \leq \frac{1}{a^2}\Phi(x) $, the above implies that 
$$\sum_{i=1}^\infty \mu_i \Phi\pa{ \frac{f_i}{\sqrt{3}}} \leq 1 $$
and as such, by the definition of $\norm{\cdot}_{L_\Phi}$, we conclude that
$$\norm{\bm{f}}_{L_\Phi}^2 \leq 3=\frac{3}{2}\mathcal{L}(\bm{f}),$$
and the proof is complete.
\end{proof}
\begin{proof}[Proof of Theorem \ref{thm:log sobelev and poincare I}]
The equivalence of $(ii)$ and $(iii)$ is immediate following Proposition \ref{prop:poincare and L}, which also proves the desired connection between $\Lambda$ and $\lambda$. To show that $(i)$ implies $(ii)$ we notice that as the right hand side of \eqref{eq: log-sob} is invariant under translation. Taking the supremum over all possible translations results in $(ii)$. The fact that $(ii)$ implies $(i)$ is immediate as 
$$\Ent_{\mu}(\bm{f}^2) \leq \mathcal{L}(\bm{f}).$$
\end{proof}

This observation that the log-Sobolev inequality with weights is actually a form of a Poincar\'e inequality brings to mind another inequality with weights that is closely connected to the Poincar\'e inequality - Hardy inequality. In its discrete form, we have that
\begin{thm}\label{thm: Hardy inequality general V.1}
Let $\bm{\mu}$ and $\bm{\nu}$ two sequences of positive numbers and let $m\in \N$. Then, the following two conditions are equivalent:
\begin{enumerate}[(i)]
\item There exists a finite constant $A_{1,m} \geq 0$ such that 
$$\sum_{i=m}^\infty \mu_i \pa{\sum_{j=m}^i f_j}^2 \leq A_{1,m} \sum_{i=m}^\infty \nu_i f_i^2,$$
for any sequence $\bm{f}$.
\item The following holds:
$$B_{1,m}=\sup_{k\geq m}\pa{\sum_{i=k}^\infty \mu_i}\pa{\sum_{i=m}^k \frac{1}{\nu_i}}<\infty.$$
\end{enumerate}
Moreover, if any of the conditions holds than $B_{1,m}\leq A_{1,m} \leq 4B_{1,m}$.
\end{thm}
The proof for the case $m=1$ can be found in \cite{JACBL}, and the general case follows by the same method of proof.
\begin{cor}\label{cor: Hardy inequality V.1}
Let
$$B^{(1)}_{m}=\sup_{k\geq m}\pa{\sum_{i=k+1}^\infty \mu_i}\pa{\sum_{i=m}^k \frac{1}{\nu_i}}.$$
Then for any sequence $\bm{f}$ such that $f_m=0$ we have that
\begin{equation}\label{eq: Hardy inequality V.1}
\sum_{i=m}^\infty \mu_i f_{i}^2 \leq A^{(1)}_{m} \sum_{i=m}^\infty \nu_i \pa{f_{i+1}-f_i}^2,
\end{equation}
if and only if $B^{(1)}_{m}<\infty$. In that case $B^{(1)}_m\leq A^{(1)}_{m} \leq 4B^{(1)}_m$. Additionally, 
$$B_{1,m} \leq B^{(1)}_m \leq B_{1,m+1}.$$
\end{cor}
\begin{proof}
This follows immediately form Theorem \ref{thm: Hardy inequality general V.1} applied to the sequence $g_i=f_{i+1}-f_i$ and a simple translation argument.
\end{proof}
Besides the above, we will also need to have a Hardy-type inequality for sums up to a fixed integer $m$. 
\begin{thm}\label{thm: Hardy inequality general V.2}
Let $\bm{\mu}$ and $\bm{\nu}$ two sequences of positive numbers and let $m\in \N$. Then, for any sequence $\bm{f}$ such that $f_m=0$ we have that if there exists $A>0$ such that
\begin{equation}\label{eq: Hardy inequality general V.2}
\sum_{i=1}^{m-1} \mu_i f_i^2 \leq A \sum_{i=1}^{m-1} \nu_i \pa{f_{i+1}-f_i}^2,
\end{equation}
then $ b_{2,m}\leq A $ where
$$b_{2,m}=\sup_{k\leq m-1}\sum_{i=1}^{k} \mu_i \pa{\sum_{j=k}^{m-1} \frac{1}{\nu_j}}.$$
Moreover, one can always choose
$$A=B_{2,m}=\sum_{i=1}^{m-1} \mu_i \pa{\sum_{j=i}^{m-1} \frac{1}{\nu_j}}.$$
\end{thm}
\begin{proof}
We start by noticing that for any $1\leq i \leq m-1$ we have that
\begin{equation*}\begin{split}f_i^2 = \left[\sum_{j=i}^{m-1}\pa{f_{j+1}-f_j }\right]^2 &\leq \pa{\sum_{j=i}^{m-1}\frac{1}{\nu_j}} \pa{\sum_{j=i}^{m-1}\nu_j \pa{f_{j+1}-f_j}^2}\\
&\leq \pa{\sum_{j=i}^{m-1}\frac{1}{\nu_j}} \pa{\sum_{j=1}^{m-1}\nu_j \pa{f_{j+1}-f_j}^2}.\end{split}\end{equation*}
Thus
$$\sum_{i=1}^{m-1}\mu_i f_i^2 \leq \left[\sum_{i=1}^{m-1} \mu_i \pa{ \sum_{j=i}^{m-1}\frac{1}{\nu_j}}\right] \pa{\sum_{j=1}^{m-1}\nu_j \pa{f_{j+1}-f_j}^2}=B_{2,m}\sum_{j=1}^{m-1}\nu_j \pa{f_{j+1}-f_j}^2,$$
completing the second statement. Next, for any $j \leq m-1$ we denote by
$$\sigma_j=\sum_{i=j}^{m-1}\frac{1}{\nu_i}.$$
Fix $k\leq m-1$ and define $\bm{f}^{(k)}$ to be such that $f^{(k)}_i=\sigma_k$ when $i\leq k$ and $f^{(k)}_i=\sigma_i$ when $i>k$. We have that
$$\sum_{i=1}^{m-1}\nu_i \pa{f^{(k)}_{i+1}-f^{(k)}_i}^2=\sum_{i=k}^{m-1}\nu_i \pa{f^{(k)}_{i+1}-f^{(k)}_i}^2=\sum_{i=k}^{m-1}\frac{1}{\nu_i}=\sigma_k.$$
On the other hand
$$\sum_{i=1}^{m-1} \mu_i \pa{f^{(k)}_i}^2 \geq \sum_{i=1}^{k} \mu_i \pa{f^{(k)}_i}^2=\sigma_k^2 \pa{\sum_{i=1}^{k} \mu_i }.$$
As \eqref{eq: Hardy inequality general V.2} is valid we see that $A\geq \pa{\sum_{i=k}^{m-1}\frac{1}{\nu_i}}\pa{\sum_{i=1}^{k}\mu_i}$ for all $k$. This completes the proof.
\end{proof}
As we can see, the expression for the constants $B^{(1)}_m$ and $B^{(2)}_m$ are starting to look similar to the expression appearing in $(ii)$ of Theorem \ref{thm: log sobolev condition}. However, we still need a few more technicalities to complete the proof.
\begin{thm}\label{thm: log sobolev equivalent} 
The following conditions are equivalent:
\begin{enumerate}[(i)]
\item  $\bm{\nu}$ admits a log-Sobolev inequality with respect to $\bm{\mu}$ with constant $\Lambda$.
\item There exists $\eta > 0$ such that, for any sequence $\bm{f}=\br{f_i}$ such that $f_m=0$ with $m\in\N$ satisfying
$$\max\pa{\sum_{i=1}^{m-1} \mu_i , \sum_{i=m+1}^\infty \mu_i}<\frac{2}{3}$$
we have that
$$\norm{\pa{\bm{f}^{(0)}}^2}_{L_\Psi}+\norm{\pa{\bm{f}^{(1)}}^2}_{L_\Psi} \leq \eta \sum_{i=1}^\infty \nu_i \pa{f_{i+1}-f_i}^2,$$
where $\bm{f}^{(0)}=\bm{f} 1\!\!1_{i<m}$ and $\bm{f}^{(1)}=\bm{f} 1\!\!1_{i>m}$.
\end{enumerate}
Moreover, if condition $(ii)$ is valid one can choose $\Lambda=40\eta$.
\end{thm}
\begin{proof}
Using Theorem \ref{thm:log sobelev and poincare I} we notice that it is enough for us to show the equivalence of conditions $(ii)$ of our theorem and that of Theorem \ref{thm:log sobelev and poincare I}.

Assume, to begin with, that $(ii)$ of Theorem \ref{thm:log sobelev and poincare I} is valid. As was shown in the aforementioned theorem, this implies that 
\begin{equation}\label{eq:log sobolev equivalence proof I}
\norm{\bm{f}-\avg{\bm{f}}}^2_{L_\Phi} \leq \frac{3\Lambda}{2}  \sum_{i=1}^\infty \nu_i \pa{f_{i+1}-f_i}^2.
\end{equation}
Due to the conditions on $\bm{f}$ and the definition of $\bm{f}^{(0)}$ and $\bm{f}^{(1)}$ one has that
\begin{equation*}
\begin{split}
\norm{\avg{\bm{f}^{(0)}}}_{L_\Phi } &\leq \abs{\avg{\bm{f}^{(0)}}} \leq \norm{\bm{f}^{(0)}}_{L^2_\mu}\sqrt{\sum_{i=1}^{m-1} \mu_i}\\
\norm{\avg{\bm{f}^{(1)}}}_{L_\Phi } &\leq \abs{\avg{\bm{f}^{(1)}}} \leq \norm{\bm{f}^{(1)}}_{L^2_\mu}\sqrt{\sum_{i=m+1}^\infty \mu_i}
\end{split}
\end{equation*}
(see Lemma \ref{lemapp: evaluation of norm of c^0 and c^1} in Appendix \ref{app:discrete log-sob}). Thus
$$
\norm{\bm{f}^{(0)}}_{L_\Phi} \leq \norm{\bm{f}^{(0)}-\avg{\bm{f}^{(0)}}}_{L_\Phi}+\norm{\avg{\bm{f}^{(0)}}}_{L_\Phi} \leq \norm{\bm{f}^{(0)}-\avg{\bm{f}^{(0)}}}_{L_\Phi} 
+ \sqrt{\frac{3}{2}\sum_{i=1}^{m-1}\mu_i}\norm{\bm{f}^{(0)}}_{L_\Phi},$$
implying that
$$\norm{\bm{f}^{(0)}}_{L_\Phi} \leq \frac{1}{1-\sqrt{\frac{3}{2}\sum_{i=1}^{m-1}\mu_i}}\norm{\bm{f}^{(0)}-\avg{\bm{f}^{(0)}}}_{L_\Phi},$$
and similarly
$$\norm{\bm{f}^{(1)}}_{L_\Phi} \leq \frac{1}{1-\sqrt{\frac{3}{2}\sum_{i=m+1}^{\infty}\mu_i}}\norm{\bm{f}^{(1)}-\avg{\bm{f}^{(1)}}}_{L_\Phi}.$$
We can conclude, by applying (\ref{eq:log sobolev equivalence proof I}) to $\bm{f}^{(0)}$ and $\bm{f}^{(1)}$, that
\begin{multline*}
\norm{\bm{f}^{(0)}}^2_{L_\Phi}  \leq \frac{3\Lambda}{2\pa{1-\sqrt{\frac{3}{2}\sum_{i=1}^{m-1}\mu_i}}^2} \sum_{i=1}^{m-1}\nu_i \pa{f_{i+1}-f_i}^2\\
{\text{ and }\qquad  \norm{\bm{f}^{(1)}}^2_{L_\Phi}  \leq \frac{3\Lambda}{2\pa{1-\sqrt{\frac{3}{2}\sum_{i=m+1}^{\infty}\mu_i}}^2} \sum_{i=m}^{\infty}\nu_i \pa{f_{i+1}-f_i}^2.} \end{multline*}
The result now follows from \eqref{eq: observation about L_Phi and L_Psi}.

To show the converse, we use the translation invariance of $(ii)$ from Theorem  \ref{thm:log sobelev and poincare I} to assume that $f_m=0$. As such we have that $\bm{f}=\bm{f}^{(0)}+\bm{f}^{(1)}$. Moreover,
\begin{multline*}
\norm{\bm{f}-\avg{\bm{f}}}^2_{L_\Phi}\leq \pa{\norm{\bm{f}^{(0)}-\avg{\bm{f}^{(0)}}}_{L_\Phi}+\norm{\bm{f}^{(1)}-\avg{\bm{f}^{(1)}}}_{L_\Phi}}^2\\
\leq \pa{\pa{1+\sqrt{\frac{3}{2}}\sqrt{\sum_{i=1}^{m-1}\mu_i}}\norm{\bm{f}^{(0)}}_{L_\Phi}+\pa{1+\sqrt{\frac{3}{2}}\sqrt{\sum_{i=m+1}^{\infty}\mu_i}}\norm{\bm{f}^{(1)}}_{L_\Phi}}^2\\
\leq 2\pa{1+\sqrt{\frac{3}{2}}\sqrt{\sum_{i=1}^{m-1}\mu_i}}^2\norm{\bm{f}^{(0)}}_{L_\Phi}^2+2\pa{1+\sqrt{\frac{3}{2}}\sqrt{\sum_{i=m+1}^{\infty}\mu_i}}^2\norm{\bm{f}^{(1)}}^2_{L_\Phi}\\
\leq 2\eta\max\pa{\pa{1+\sqrt{\frac{3}{2}}\sqrt{\sum_{i=1}^{m-1}\mu_i}}^2,\pa{1+\sqrt{\frac{3}{2}}\sqrt{\sum_{i=m+1}^{\infty}\mu_i}}^2}  \sum_{i=1}^\infty \nu_i \pa{f_{i+1}-f_i}^2\end{multline*}
where we again used (\ref{eq: observation about L_Phi and L_Psi}). This shows the desired result due to Theorem \ref{thm:log sobelev and poincare I}.
\end{proof}
We have finally gained all the tools we need to prove Theorem \ref{thm: log sobolev condition}.
\begin{proof}[Proof of Theorem \ref{thm: log sobolev condition}]
  Our main tool will be Theorem \ref{thm: log sobolev equivalent}. It is known that
  \begin{equation}\nonumber
\norm{\bm{f}^2}_{L_\Psi} = \sup \br{\sum_{i=1}^\infty \mu_i f^2_i g_i \; ; \; \sum_{i=1}^\infty \mu_i \Xi(g_i) \leq 1},
\end{equation}
where $\Xi$ is the Young complement of $\Psi$. Using Corollary \ref{cor: Hardy inequality V.1} we know that if $f_m=0$ then
$$\sum_{i=m}^\infty \mu_if^2_i g_i \leq \Lambda \sum_{i=m}^\infty \nu_i \pa{f_{i+1}-f_i}^2 $$
if and only if 
$$B=\sup_{k\geq m}\pa{\sum_{i=k+1}^\infty g_i \mu_i}\pa{\sum_{i=1}^k \frac{1}{\nu_i}} <\infty.$$
Taking supremum over all appropriate $\bm{g}=\br{g_{i}}$, we find that
\begin{equation}\label{eq: log sobelev condition 1.5}
\norm{\bm{f}^2 1\!\!1_{i>m}}_{L_\Psi} \leq \Lambda \sum_{i=m}^\infty \nu_i \pa{f_{i+1}-f_i}^2  
\end{equation}
if and only if
$$B=\sup_{k\geq m}{\norm{1\!\!1_{[k+1,\infty)}}_{L_\Psi}}{\sum_{i=1}^k \frac{1}{\nu_i}} <\infty.$$
As
\begin{equation*}\begin{split}\norm{1\!\!1_{[k+1,\infty)}}_{L_\Psi}&=\inf_{\alpha>0}\br{\sum_{i=k+1}^\infty \mu_i \Psi\pa{\frac{1}{\alpha}} \leq 1}=\inf_{\alpha>0}\br{\Psi\pa{\frac{1}{\alpha}}\leq \frac{1}{\sum_{i=k+1}^\infty \mu_i}}\\
&=\frac{1}{\Psi^{-1}\pa{\frac{1}{\sum_{i=k+1}^\infty \mu_i}}}\end{split}\end{equation*}
we find that \eqref{eq: log sobelev condition 1.5} is equivalent to $B_1<\infty$, showing that $(i)$ implies $(ii)$. \\
Conversely, using Theorem \ref{thm: Hardy inequality general V.2} we find that if $f_m=0$ then
\begin{equation*}\begin{split}\sum_{i=1}^{m-1} \mu_i f^2_i g_i &\leq  \left[\sum_{i=1}^{m-1} \mu_i g_i \pa{\sum_{j=i}^{m-1}\frac{1}{\nu_j}} \right]\sum_{i=1}^{m-1}\nu_i\pa{f_{i+1}-f_i}^2 \\
&\leq \left[\pa{\sum_{i=1}^{m-1} \mu_i g_i }\pa{\sum_{j=1}^{m-1}\frac{1}{\nu_j}}\right]\sum_{i=1}^{m-1}\nu_i\pa{f_{i+1}-f_i}^2\end{split}\end{equation*}and again, by taking supremum over the appropriate $\bm{g}$, we find that
\begin{equation}\label{eq: log sobelev condition 1.75}
\norm{\bm{f}^2 1\!\!1_{i<m}}_{L_\Psi} \leq B_2 \sum_{i=1}^{m-1} \nu_i \pa{f_{i+1}-f_i}^2.  
\end{equation}
Thus, if $\bm{f}=\br{f_i}$ is a sequence such that $f_m=0$, and if in addition $B_1<\infty$ we have that
\begin{equation*}\begin{split}\norm{\pa{\bm{f}^{(0)}}^2}_{L_\Psi}+\norm{\pa{\bm{f}^{(1)}}^2}_{L_\Psi} &\leq  B_2 \sum_{i=1}^{m-1} \nu_i \pa{f_{i+1}-f_i}^2+ 4 B_1 \sum_{i=m}^{\infty} \nu_i \pa{f_{i+1}-f_i}^2\\
&\leq \pa{B_2+4B_1}\sum_{i=1}^\infty \nu_i \pa{f_{i+1}-f_i}^2,\end{split}\end{equation*} 
where we have used Corollary \ref{cor: Hardy inequality V.1}. We conclude, using Theorem \ref{thm: log sobolev equivalent}, that if $B_1<\infty$ then $\bm{\nu}$ admits a log-Sobolev inequality with respect to $\bm{\mu}$ with constant $\Lambda$ that can be chosen to be $\Lambda=40(B_1+4B_2).$
\end{proof}
We are only left with the proof of Corollary \ref{cor: log sobolev equivalent}. The proof relies on the following technical lemma, whose proof is left to Appendix \ref{app:discrete log-sob}:
\begin{lem}\label{lem: evaluation of Psi inverse }
For any $t\geq \frac{3}{2}$ one has that
\begin{equation*}
\frac{1}{3}\frac{t}{\log t} \leq \Psi^{-1}(t) \leq 2 \frac{t}{\log t}.
\end{equation*}
\end{lem}

\begin{proof}[Proof of Corollary \ref{cor: log sobolev equivalent}]
Due to the choice of $m$ and Lemma \ref{lem: evaluation of Psi inverse } we know that $\Psi^{-1}(t)$ and $\frac{t}{\log t}$ are equivalent for our choice of
$$t=\frac{1}{\sum_{i=m+1}^\infty \mu_i}.$$
This shows the desired equivalence using Theorem \ref{thm: log sobolev condition}. As for the last estimation, it follows immediately from the fact that 
$$B_i \leq 3 D_i,$$
for $i=1,2$. 
\end{proof}

Now that we have achieved a necessary and sufficient condition to the validity of a discrete log-Sobolev inequality with weight, we will proceed to see how it can be used to prove Theorem \ref{thm:main thm fe}.

\section{Energy Dissipation Inequalities}
\label{sec:EDI}

\subsection{Cercignani's Conjecture for the Becker-D\"oring equations}

Motivated by our previous section, the first step in trying to show
the validity of Cercignani's conjecture would be to connect between
the energy dissipation, $D(\bm{c})$, and a term that resembles the
right hand side of \eqref{eq: log-sob}. Recall that, for any
non-negative sequence $\bm{c}=\br{c_{i}}$ we defined
$$D(\bm{c})=\sum_{i=1}^{\infty} a_i {Q}_i
  \pa{\frac{c_1 c_i}{{Q}_i}-\frac{c_{i+1}}{Q_{i+1}}} \pa{\log
    \pa{\frac{c_1 c_i}{Q_i}} -\log\pa{\frac{c_{i+1}}{Q_{i+1}}}}$$ and
$$\overline{D}(\bm{c})=\sum_{i=1}^\infty a_i Q_i\pa{\sqrt{\frac{c_1c_i}{Q_i}} - \sqrt{\frac{c_{i+1}}{Q_{i+1}}}}^2$$

We have the following properties:
\begin{lem}\label{lem:lower energy properties} For any non-negative sequence $\bm{c}$, the following holds
\begin{enumerate}[(i)]
\item  We have that
\begin{equation}\label{eq:lower energy property I}
\overline{D}(\bm{c}) \leq D(\bm{c})
\end{equation}
\item For any $z>0$ we have that
\begin{equation}\label{eq:lower energy property II}
\begin{gathered}
D(\bm{c})=\sum_{i=1}^\infty a_i \pa{\Q_z}_i \pa{\Q_z}_1 \pa{\frac{c_1 c_i}{\pa{\Q_z}_i \pa{\Q_z}_1}-\frac{c_{i+1}}{\pa{\Q_z}_{i+1} }}\\
\pa{\log \pa{\frac{c_1 c_i}{\pa{\Q_z}_i \pa{\Q_z}_1}}
-\log\pa{\frac{c_{i+1}}{\pa{\Q_z}_{i+1} }}} \\
\end{gathered}
\end{equation}
and
\begin{equation}\label{eq:lower energy property III}
\overline{D}(\bm{c})=\sum_{i=1}^\infty a_i \pa{\Q_z}_i \pa{\Q_z}_1 \pa{\sqrt{\frac{c_1 c_i}{\pa{\Q_z}_i \pa{\Q_z}_1}} - \sqrt{\frac{c_{i+1}}{\pa{\Q_z}_{i+1} }}}^2
\end{equation}
\end{enumerate}
\end{lem}
 \begin{proof}
 $(i)$ is an immediate consequence of the inequality
 $$(x-y)\pa{\log x -\log y} \geq \pa{\sqrt{x}-\sqrt{y}}^2$$
and $(ii)$ is immediate from the definition of $\Q_z$ and the homogeneous nature of the expressions involved.
 \end{proof}
We note that property $(ii)$ of the above lemma gives an indication of how we may be able to find a connection between $\overline{D}(\bm{c})$ and $H(\bm{c}|\Q)$. However, $\Q$ is not the only equilibrium state we need to consider. Similar to the work of Jabin and Niethammer (\cite{JN03}), another 'equilibrium' state that will play an important role in what is to follow is
 $$\tilde{\Q}=Q_{c_1}.$$
Indeed, it is the \emph{only} possible 'equilibrium' under which the right hand side of \eqref{eq:lower energy property III} attains a form that is suitable for the log-Sobolev theory we developed. From \eqref{eq:lower energy property III} we find that
\begin{equation}\label{eq:D(c) bigger than log sob related}
\begin{gathered}
\overline{D}(\bm{c})= \sum_{i=1}^\infty a_i \tilde{\Q}_i \tilde{\Q_1}\pa{\sqrt{\frac{ c_i}{\tilde{\Q}_i}} - \sqrt{\frac{c_{i+1}}{\tilde{\Q}_{i+1} }}}^2
\end{gathered}
\end{equation}
 \begin{prp}\label{prop:good things if there is log-sob}
 For given coagulation and detailed balance coefficients, $\br{a_i}_{i\in \N},\br{Q_i}_{i\in \N}$, and a given positive sequence $\bm{c}$ with finite mass $\varrho$, we define the following measures
 \begin{equation}\label{eq:definition of measures for log-sob}
  \mu_i = \frac{\tilde{\Q}_i}{\sum_{i=1}^\infty
  \tilde{\Q}_i},
  \qquad
  \nu_i := \frac{a_i \tilde{\Q}_i}{\sum_{j=1}^\infty a_j \tilde{\Q}_j},
  \qquad i \in \N.
 \end{equation}
 Then, if $\bm{\nu}$ admits a log-Sobolev inequality with respect to $\bm{\mu}$ with constant $\Lambda$ we have that
 \begin{equation}\label{eq:good things if there is log-sob}
 \overline{D}(\bm{c}) \geq \frac{ c_1^3 \pa{\sum_{i=1}^\infty a_{i}\tilde{\Q}_{i}}}{\Lambda \pa{\sum_{i=1}^\infty \tilde{\Q}_i }\pa{c_1^2+2\pa{\sum_{i=1}^\infty c_i}\pa{\sum_{i=1}^\infty \tilde{\Q}_i }}}H(\bm{c}|\Q)
\end{equation}
 \end{prp}
\begin{proof}
Denote by $f_i=\sqrt{\frac{c_i}{\tilde{\Q}_i}}$. Because $\bm{\nu}$ admits a log-Sobolev inequality with respect to $\bm{\mu}$ with constant $\Lambda$ we have that
\begin{equation}\label{eq:prop good things proof I}
\begin{gathered}
{\overline{D}(\bm{c})=}\pa{\sum_{i=1}^\infty a_i \tilde{\Q}_i \tilde{\Q}_1}\sum_{i=1}^\infty \nu_i \pa{f_{i+1}-f_i}^2 \geq \frac{c_1  \pa{\sum_{i=1}^\infty a_i \tilde{\Q}_i }}{\Lambda}\Ent_{\mu}\pa{\bm{f}^2}.
\end{gathered}
\end{equation}
Next, we notice that
\begin{multline}
  \label{eq:2}
  \left( \sum_{i=1}^\infty \tilde{\Q}_i \right)
  \Ent_\mu(\bm{f}^2)
  =
  \sum_{i=1}^\infty c_i \log
  \frac{c_i}{\tilde{\Q}_i}
  - \left( \sum_{i=1}^\infty c_i \right)
  \left( \log \sum_{i=1}^\infty c_i - \log \sum_{i=1}^\infty \tilde{\Q}_i \right)
  \\
  =
  H(\bm{c}|\tilde{\Q})
  + \sum_{i=1}^\infty c_i
  - \sum_{i=1}^\infty \tilde{\Q}_i
  - \left( \sum_{i=1}^\infty c_i \right)
  \left( \log \sum_{i=1}^\infty c_i - \log \sum_{i=1}^\infty \tilde{\Q}_i \right)
  \\
  =
  H(\bm{c}|\tilde{\Q})
  -
  \left( \sum_{i=1}^\infty \tilde{\Q}_i \right)
  \Delta
  \left(
    \frac{\sum_{i=1}^\infty c_i}{\sum_{i=1}^\infty \tilde{\Q}_i}
  \right),
\end{multline}
where $\Delta(x)=x\log x -x+1$. Using the fact that
\begin{equation}
  \label{eq:relative-entropy-min}
  H(\bm{c} | \tilde{\Q}) \geq H(\bm{c} | \Q)
\end{equation}
(see Lemma \ref{lemapp:H(c|Q) minimizes } in Appendix \ref{app:additional useful}) and the following Csiszár-Kullback inequality
\begin{equation}
  \label{eq:CK}
  \Ent_\mu(\bm{f}^2) \geq \frac{1}{2\avg{\bm{f}^2}}
  \left( \sum_{i=1}^\infty |f_i^2 - \avg{\bm{f}^2}| \mu_i \right)^2,
\end{equation}
where
\begin{equation*}
  \avg{\bm{f}^2} := \sum_{i=1}^\infty f_i^2 \mu_i.
\end{equation*}
we find that in our particular setting
\begin{equation*}
\begin{split}
  \Ent_\mu(\bm{f}^2) &\geq \frac{\sum_{i=1}^\infty \tilde{\Q}_i}{2\sum_{i=1}^\infty c_i}\pa{\sum_{i=1}^\infty \abs{\frac{c_i}{\sum_{i=1}^\infty \tilde{\Q}_i}-\frac{\tilde{\Q}_i\pa{\sum_{i=1}^\infty c_i}}{\pa{\sum_{i=1}^\infty \tilde{\Q}_i}^2}}}^2 \\
  &=\frac {\sum_{i=1}^\infty c_i}{2\sum_{i=1}^\infty \tilde{\Q}}_i
  \left( \sum_{i=1}^\infty \left| \frac{c_i}{\sum_{i=1}^\infty c_i}
      - \frac{\tilde{\Q}_i}{\sum_{i=1}^\infty \tilde{\Q}_i}
    \right|
  \right)^2 \end{split}
\end{equation*}
and keeping only the first term in the last sum we get
\begin{equation*} \Ent_{\mu}(\bm{f}^{2})\geq \frac {\sum_{i=1}^\infty c_i}{2\sum_{i=1}^\infty \tilde{\Q}}_i
  \left(  \left| \frac{c_1}{\sum_{i=1}^\infty c_i}
      - \frac{\tilde{\Q}_1}{\sum_{i=1}^\infty \tilde{\Q}_i}
    \right|
  \right)^2 
=\frac{c_1^2}{2\sum_{i=1}^\infty c_i \sum_{i=1}^\infty \tilde{\Q_i}}
  \left( 1 - \frac{\sum_{i=1}^\infty c_i}{ \sum_{i=1}^\infty \tilde{\Q_i}} \right)^2
\end{equation*}
Continuing from \eqref{eq:2} and using \eqref{eq:relative-entropy-min}, the above inequality and the fact that
$$\Delta(x) \leq (x-1)^2$$
we find that 
\begin{multline*}
  \left( \sum_{i=1}^\infty \tilde{\Q_i} \right) \Ent_\mu \pa{\bm{f}^2}
  \geq
  H(\bm{c} | \Q) - \left( \sum_{i=1}^\infty \tilde{\Q_i} \right)
  \left(
    \frac{\sum_{i=1}^\infty c_i}{ \sum_{i=1}^\infty \tilde{\Q_i}} - 1
  \right)^2
  \\
  \geq
  H(\bm{c} | \Q) - \frac{2}{c_1^2}
  \left( \sum_{i=1}^\infty \tilde{\Q_i} \right)^2 \left( \sum_{i=1}^\infty c_i \right)
  \Ent_\mu \pa{\bm{f}^2}.
\end{multline*}
Thus,
\begin{equation*}
  H(\bm{c} | \Q)
  \leq
  \left( \sum_{i=1}^\infty \tilde{\Q_i} \right)
  \left( 1 + \frac{2}{c_1^2}
    \left( \sum_{i=1}^\infty \tilde{\Q_i} \right) \left( \sum_{i=1}^\infty c_i \right)
  \right)
  \Ent_\mu \pa{\bm{f}^2}	.
\end{equation*}
Combining the above with \eqref{eq:prop good things proof I} completes the proof.
\end{proof} 
\begin{cor}\label{cor:cor good things if there is log-sob}
Under the conditions of Proposition \ref{prop:good things if there is log-sob} and the additional assumption that $\varrho<\varrho_s<+\infty$  and $c_1<z_s$ we have that
\begin{equation}\label{eq:cor good things if there is log-sob}
 {\overline{D}(\bm{c}) }\geq \frac{ a_1z_s^2 c_1^2 }{\Lambda \pa{z_s+\varrho_s}\pa{z_s^2+2\varrho(z_s+\varrho_s)}}H(\bm{c}|\Q)
\end{equation}
\end{cor} 
 \begin{proof}
 This follows immediately from \eqref{eq:good things if there is log-sob} and the following estimates 
$$\sum_{i=1}^\infty \tilde{\Q}_i = \sum_{i=1}^\infty Q_i c_1^i \leq c_1\pa{1+\frac{1}{z_s}\sum_{i=2}^\infty Q_i z_s^i} <c_1\pa{1+\frac{\varrho_s}{z_s}}$$
$$\sum_{i=1}^\infty c_i \leq \sum_{i=1}^\infty i c_i =\varrho$$
together with $\sum_{i=1}^{\infty}a_{i}\tilde{\Q}_{i} \geq a_{1}c_{1}.$
 \end{proof}
 Corollary \ref{cor:cor good things if there is log-sob} shows us that as long as $c_1$ is bounded from below, Cercignani's conjecture will follow immediately form a log-Sobolev inequality for $\bm{\nu}$ with respect to $\bm{\mu}$, defined in Proposition \ref{prop:good things if there is log-sob}. Our next Proposition will show that it is true, in a specific setting.
 \begin{prp}\label{prop:log-sob is valid if gamma=1}
 Let $\br{a_i}_{i\in\N},\br{Q_i}_{i\in\N}$ satisfy Hypothesis \ref{hyp: z_s>0}-\ref{hyp: condition on a_i} with $\gamma=1$ and let $\bm{c}=\br{c_i}_{i\in\N}$ be an arbitrary positive sequence with finite total density $\varrho<\varrho_s<+\infty$. Assume that there exists $\delta>0$ such that
\begin{equation*}
c_1 \leq \overline{z}+\delta <z_s.
\end{equation*}
Then, the measure $\bm{\nu}$ admits a log-Sobolev inequality with respect to the measure $\bm{\mu}$ with constant
\begin{equation}\label{eq:log-sob is valid if gamma=1}
\begin{gathered}
\Lambda=120\frac{C\pa{\frac{\overline{z}+\delta}{z_s}}}{e}\frac{z_s^3 }{(z_s-\overline{z}-\delta)^3}\Bigg(\frac{3(\overline{z}+\delta)}{z_s}\\
+ \pa{1+\frac{2e(\overline{z}+\delta)}{z_s}\sup_k\abs{\log \pa{\alpha_{k+1}^{\frac{1}{k+1}}}} + \frac{e(\overline{z}+\delta)}{z_s} \log\pa{\frac{z_s}{z_s-\overline{z}-\delta}}}\Bigg)\\
\end{gathered}
\end{equation}
where $\bm{\mu}$ and $\bm{\nu}$ were defined in Proposition \ref{prop:good things if there is log-sob} and
$$C(\eta)=1+ \sup_{k\geq 3} \pa{k\pa{1+\log \pa{\frac{k}{2}}}\eta_1^{\frac{k}{2}}}+\frac{2\eta}{1-\eta},$$
for $\eta<1$. 
 \end{prp}
\begin{proof}
As mentioned in the introduction to our work, we can assume without loss of generality that $a_i=i$. We may also assume that $\alpha_1$, from Hypothesis \ref{hyp: Qi-form}, equals $1$. We denote by 
$$\eta=\frac{c_1}{z_s}\leq \frac{\overline{z}+\delta}{z_s}=\eta_1<1.$$
As
$$\tilde{\Q}_i = \alpha_i z_s^{1-i}c_1^i = z_s\alpha_i \eta^i$$
we find that due to the monotonicity of $\br{\alpha_i}_{i\in\N}$
$$ z_s\alpha_{k+1}\eta^{k+1}=\tilde{\Q}_{k+1}\leq \sum_{i=k+1}^\infty \tilde{\Q}_i \leq  z_s \eta^{k+1}\sum_{i=1}^\infty \alpha_{i+k}\eta^{i-1} \leq \frac{z_s \alpha_{k+1}\eta^{k+1}}{1-\eta}.$$
As such 
$$ \alpha_{k+1}(1-\eta)\eta^k \leq\sum_{i=k+1}^\infty \mu_i \leq \alpha_{k+1}  \frac{\eta^k}{1-\eta}, $$
implying that
\begin{equation}\label{eq:log-sob is valid if gamma=1 proof I}
-\sum_{i=k+1}^\infty \mu_i \log \pa{\sum_{i=k+1}^\infty \mu_i} \leq \frac{\alpha_{k+1}\eta^{k}}{1-\eta}\pa{k\log \pa{\frac{1}{\eta}} -\log\pa{\alpha_{k+1}(1-\eta)}}.
\end{equation}
Next, we notice that as 
$$\sum_{i=1}^\infty i y^i = \frac{y}{(1-y)^2},$$
one has that
$$ z_s \eta \leq \sum_{i=1}^\infty i\alpha_i z_s \eta^i = \sum_{i=1}^\infty a_i \tilde{\Q}_i \leq  z_s \frac{\eta}{(1-\eta)^2}, $$
from which we find that
$$i\alpha_i (1-\eta)^2\eta^{i-1} \leq \nu_i \leq i\alpha_i\eta^{i-1}.$$
We notice that for $k\geq 3$ the monotonicity of $\br{\alpha_i}_{i\in \N}$ implies that
$$k\alpha_k \eta^k \sum_{i=1}^k \frac{1}{i\alpha_i}\pa{\frac{1}{\eta}}^i = 1 +\sum_{i=1}^{k-1}\frac{k\alpha_k}{i\alpha_i}\eta^{k-i}$$
$$ \leq 1+ \sum_{i=1}^{k-1}\frac{k}{i}\eta^{k-i} = 1+ \sum_{i=1}^{\left[ \frac{k}{2}\right]}\frac{k}{i}\eta^{k-i} + \sum_{i=\left[ \frac{k}{2}\right]+1}^{k-1}\frac{k}{i}\eta^{k-i}  \leq 1+ k \eta_1^{\frac{k}{2}}\sum_{i=1}^{\left[ \frac{k}{2}\right]}\frac{1}{i}+\frac{k}{\left[ \frac{k}{2}\right]+1}\sum_{j=1}^\infty \eta_1^j$$
$$\leq 1+ k\pa{1+\log \pa{\frac{k}{2}}}\eta_1^{\frac{k}{2}}+\frac{2\eta_1}{1-\eta_1}.$$
Using the definition of $C(\eta)$ and the fact that $C(\eta)>1+\eta$ we find that for all $k\in\N$
$$k\alpha_k \eta^k \sum_{i=1}^k \frac{1}{i\alpha_i}\pa{\frac{1}{\eta}}^i  \leq C(\eta_1).$$
and as such
\begin{equation}\label{eq:sum of 1/nu}
\sum_{i=1}^k \frac{1}{\nu_i} \leq C(\eta_1) \frac{ \eta}{(1-\eta)^2} \frac{1}{k\alpha_k}\pa{\frac{1}{\eta}}^k
\end{equation}
Combining the above with \eqref{eq:log-sob is valid if gamma=1 proof I} yields the bound
$$\pa{-\sum_{i=k+1}^\infty \mu_i \log \pa{\sum_{i=k+1}^\infty \mu_i} }\pa{\sum_{i=1}^k \frac{1}{\nu_i} }$$
$$\leq C(\eta_1) \frac{\alpha_{k+1}}{\alpha_k}\frac{\eta}{(1-\eta)^3}\pa{\log \pa{\frac{1}{\eta}} -\frac{1}{k}\log\pa{\alpha_{k+1}(1-\eta)}}.$$
Thus, with the notation of Corollary \ref{cor: log sobolev equivalent}
\begin{equation}\label{eq:D_1 estimation}\nonumber
\begin{gathered}
D_1 \leq \frac{C(\eta_1)}{(1-\eta_1)^3} \pa{\sup_{0\leq x\leq 1}\pa{-\eta\log \pa{\eta}} +\eta_1\sup_k \frac{k+1}{k}\abs{\log \pa{\alpha_{k+1}^{\frac{1}{k+1}}}} + \eta_1\log\pa{\frac{1}{1-\eta_1}}} \\
\leq \frac{C(\eta_1)}{(1-\eta_1)^3} \pa{\frac{1}{e} +2\eta_1\sup_k\abs{\log \pa{\alpha_{k+1}^{\frac{1}{k+1}}}} + \eta_1 \log\pa{\frac{1}{1-\eta_1}}},
\end{gathered}
\end{equation}
As $m$, defined in Corollary \ref{cor: log sobolev equivalent}, is always finite we conclude using the same  Corollary that $\bm{\nu}$ admits a log-Sobolev inequality with respect to $\bm{\mu}$. However, in order to estimate the constant $\Lambda$ we still need to estimate the constant $D_2$ in the case where $m>1$ (otherwise, $D_2=0$).

Since
$$\sum_{i=m}^\infty \mu_i \leq \frac{\alpha_m}{1-\eta}\eta^{m-1}$$
the requirement that $\sum_{i=1}^{m-1}\mu_i < \frac{2}{3}$ implies that
$$\frac{1}{\alpha_{m-1} \eta^{m-1}} \leq \frac{\alpha_{m}}{\alpha_{m-1}}\frac{3}{(1-\eta)} \leq \frac{3}{(1-\eta)}.$$
Using the above along with the fact that $m>1$ and inequality \eqref{eq:sum of 1/nu} shows that
$$\sum_{i=1}^{m-1} \frac{1}{\nu_i} \leq 3 C(\eta_1)\frac{\eta_1}{(1-\eta_1)^3}\frac{1}{m-1}\leq 3 C(\eta_1)\frac{\eta_1}{(1-\eta_1)^3}.$$
We can conclude that
\begin{equation}\label{eq:log-sob is valid if gamma=1 proof II}
\begin{gathered}
\pa{-\sum_{i=m-1}^\infty \mu_i \log \pa{\sum_{i=m-1}^\infty \mu_i}}\pa{\sum_{i=1}^{m-1} \frac{1}{\nu_i}} 
\leq  3\sup_{0\leq x \leq 1}\pa{-x\log x}C(\eta_1)\frac{\eta_1}{(1-\eta_1)^3}
\end{gathered}
\end{equation}
from which we conclude that 
$$D_2 \leq \frac{3}{e}C(\eta_1)\frac{\eta_1}{(1-\eta_1)^3}$$
which completes the proof, as the result follows directly from Corollary \ref{cor: log sobolev equivalent}.
\end{proof} 
 
We finally have all the tools to prove part \eqref{enm: gamma=1 cerc conjecture} of Theorem \ref{thm:main thm fe}:
\begin{proof}[Proof of part \eqref{enm: gamma=1 cerc conjecture} of Theorem \ref{thm:main thm fe}]
The result follows immediately from Corollary \ref{cor:cor good things if there is log-sob}, Proposition \ref{prop:log-sob is valid if gamma=1} and condition \eqref{eq:good c_1}.
\end{proof} 
The last part of this  section will be devoted to the proof of part \eqref{enm: gamma<1 almost cerc conjecture} of Theorem \ref{thm:main thm fe}. For that we will need the following lemma:
\begin{lem}\label{lem:D-upper-bound}
  For any $\beta \geq 0$, any non-negative sequence $\bm{c}$ and positive sequence $\{Q_i\}_{i \geq 1}$ it holds that
  \begin{equation}
    \label{eq:D-upper-bound}
    \sum_{i=1}^\infty i^\beta Q_i  \left(\sqrt{\frac{c_1c_{i}}{Q_i}}
      - \sqrt{\frac{c_{i+1}}{Q_{i+1}}} \right)^2 \leq  2\left(c_1+ \sup_{j} \frac{Q_j}{Q_{j+1}} \right)
    \sum_{i=1}^\infty i^\beta c_i.
  \end{equation}
\end{lem} 
 \begin{proof}
  The proof is a direct result of the inequality $(a+b)^2 \leq2(a^2 + b^2)$:
  \begin{align*}
    \sum_{i=1}^\infty i^\beta Q_i
    \left(\sqrt{\frac{c_1 c_i}{Q_i}}- \sqrt{\frac{c_{i+1}}{Q_{i+1}}} \right)^2
    &\leq 2 c_1  \sum_{i=1}^\infty i^\beta c_i +2  \sum_{i=1}^\infty i^\beta \frac{Q_i}{Q_{i+1}}c_{i+i}\\
    &\leq 2\left(c_1+\sup_{j} \frac{ Q_j}{Q_{j+1}}\right)\sum_{i=1}^\infty i^\beta c_i.
  \end{align*}
\end{proof}

\begin{proof}[Proof of part \eqref{enm: gamma<1 almost cerc conjecture} of Theorem \ref{thm:main thm fe}]
We denote by $\overline{D}_\gamma (\bm{c})$ the lower free energy {dissipation} of $\bm{c}$ associated to the coagulation coefficient $a_i=i^\gamma$.   
According to part \eqref{enm: gamma=1 cerc conjecture} of Theorem \ref{thm:main thm fe}, there exists $K>0$ that depends only on $\delta,\zs,\varrho$ and $\br{\alpha_i}_{i\in \N}$   such that
$$\overline{D}_1(\bm{c})\geq K H(\bm{c}|\Q).$$
Using interpolation between $\gamma$ and $\beta$ we find that 
\begin{equation}
  \label{eq:interpolation for gamma<1}
  \overline{D}_1(c)
  \leq
  \overline{D}_\gamma^{\frac{\beta-1}{\beta-\gamma}}(c)\overline{D}_\beta^{\frac{1-\gamma}{\beta-\gamma}}(c) \leq 2^{\frac{1-\gamma}{\beta-\gamma}}\overline{D}_\gamma^{\frac{\beta-1}{\beta-\gamma}}(c) \pa{z_s+\frac{1}{z_s}\sup_j \frac{\alpha_j}{\alpha_{j+1}}}^{\frac{1-\gamma}{\beta-\gamma}}M_\beta^{\frac{1-\gamma}{\beta-\gamma}}
\end{equation}
where we have used Lemma \ref{lem:D-upper-bound}, the upper bound on
$c_1$ and Hypothesis \eqref{hyp: Qi-form}. Therefore
\begin{equation}
  \label{eq:almost-cercignani}
  D(\bm{c})
  \geq
  \overline{D}_\gamma(\bm{c}) \geq \pa{\frac{z_s
      K^{\frac{\beta-\gamma}{1-\gamma}}}{2\pa{z_s^2+\sup_j
        \frac{\alpha_j}{\alpha_{j+1}}}M_\beta}}^{\frac{1-\gamma}{\beta-1}}
  H(\bm{c}|\Q)^{\frac{\beta-\gamma}{\beta-1}}
\end{equation}
and the proof is now complete.
\end{proof}

This concludes the part of the proof of Theorem \ref{thm:main thm fe}
that relied on the log-Sobolev inequality. In the next subsection we
will address the question of what happens when $c_1$ escapes its 'good
region', given by \eqref{eq:good c_1}.

\subsection{Energy Dissipation Estimate when $c_1$ is 'Far' From
  Equilibrium }\label{sec: energy dissipation far from equilibrium}

The goal of this subsection is to show that when $c_1$ is far from equilibrium, in the aforementioned sense, then while we may lose our desired inequality between $\overline{D}(\bm{c})$ and $H(\bm{c}|\Q)$, the energy dissipation becomes \emph{uniformly large} - forcing the free energy to decrease (and as a consequence, the distance between $c_1$  and $\overline{z}$ decreases as well).

The next proposition, dealing with the case when $c_1$ is 'too large', is an adaptation of a theorem from \cite{JN03}.

\begin{prp}\label{prp:uniform lower bound on D if c_1 is large}
Let $\br{a_i}_{i\in\N},\br{Q_i}_{i\in\N}$ be the coagulation and detailed balance coefficients for the Becker-D\"oring equations. Assume that $\inf_{i}a_i >0$ and 
$$\lim_{i\rightarrow\infty}\frac{Q_{i+1}}{Q_i}=\frac{1}{z_s}.$$
Let $\bm{c}=\br{c_{i}}$ be a non-negative sequence with finite total density $\varrho<\varrho_s$. Then, if 
$$c_1 > \overline{z}+\delta$$
for any $\delta>0$, we have that
$$ {\overline{D}(\bm{c})} > \varepsilon_1,$$
for a fixed constant $\varepsilon_1$ that depends only on $\br{Q_i}_{i\in \N},\overline{z},z_s$ and $\delta$.  
\end{prp}
\begin{proof}
Without loss of generality we may assume that $\overline{z}+\delta <z_s$. Denoting by $u_i=\frac{c_i}{Q_i}$ we notice that
$${\overline{D}(\bm{c})=\sum_{i=1}^{\infty}a_{i}Q_{i}\left(\sqrt{c_{1}\,u_{i}}-\sqrt{u_{i+1}}\right)^{2}.}$$
Let $\lambda<1$ be such that $\lambda c_1 = \overline{z}+\frac{\delta}{2}$ and let $i_0\in \N$ be the first index such that
$$u_{i+1} < \lambda c_1 u_i.$$
This index exists, else, for any $i\in \N$ we have
\begin{equation}\label{eq:large c_1 proof I}
u_{i+1} \geq \lambda c_1 u_i \geq \pa{\lambda c_1}^{i} c_1
\end{equation}
and thus
$$\varrho = \sum_{i=1}^\infty ic_i  \geq c_1+c_1 \sum_{i=2}^\infty iQ_i\pa{\lambda c_1}^{i-1} \geq \sum_{i=1}^\infty iQ_i\pa{\overline{z}+\frac{\delta}{2}}^i, $$
which is a contradiction. 

Due to the positivity of each term in the sum consisting of the lower free energy dissipation, we conclude that
\begin{equation}\label{eq:Dc2}
\overline{D}(\bm{c}) \geq a_{i_0}Q_{i_0}\left(1-\sqrt{\lambda}\right)^{2}c_{1}u_{i_0}
\geq a_{i_0}Q_{i_0} \lambda^{i_0-1}c_1^{i_0+1}\left(1-\sqrt{\lambda}\right)^{2}
\end{equation}
where we have used the fact that up to $i_0-1$ we have inequality \eqref{eq:large c_1 proof I}.

As we know that there exists $C>0$, depending only on $\br{Q_i}_{i\in\N},\overline{z},z_s$ and $\delta$ such that
$$\sum_{i=i_0+1}^\infty i c_1\pa{\lambda c_1}^{i-1}Q_i \leq C  Q_{i_0}\pa{\lambda c_1}^{i_0}c_1$$
(see Lemma \ref{lemapp:control for jabin proof} in Appendix \ref{app:additional useful}), we conclude that, using \eqref{eq:large c_1 proof I} again,
$$C Q_{i_0}\pa{\lambda c_1}^{i_0}c_1 \geq \tilde{\varrho}-\sum_{i=1}^{i_0} i Q_i \pa{\lambda c_1}^{i-1}c_1 \geq \tilde{\varrho}-\sum_{i=1}^{i_0} ic_i  \geq \tilde{\varrho}-\varrho,$$
where $\tilde{\varrho}=\sum_{i=1}^\infty iQ_i \pa{\lambda c_1}^{i-1}c_1$. {We can estimate the difference $\varrho-\tilde{\varrho}$ as}
$$\tilde{\varrho}-\varrho \geq \sum_{i=1}^\infty iQ_i\pa{\pa{\overline{z}+\frac{\delta}{2}}^i-\overline{z}^i} \geq \pa{\sum_{i=1}^\infty i^2 Q_i \overline{z}^{i-1}}\frac{\delta}{2}.$$
In conclusion, there exists a universal constant $C_1>0$, depending only on $\br{Q_i}_{i\in\N},\overline{z},z_s$ and $\delta$, and not on $i_0$, $c_1$ or $\lambda$, such that
$$Q_{i_0}\pa{\lambda c_1}^{i_0}c_1 >C_1.$$
Recalling \eqref{eq:Dc2} and using the fact that $\lambda= \dfrac{\overline{z}+\frac{\delta}{2}}{c_1}<\dfrac{\overline{z}+\frac{\delta}{2}}{\overline{z}+\delta}$ we find that:
$$
\overline{D}(\bm{c}) \geq C_{1}\,a_{i_{0}}\,\dfrac{(1-\sqrt{\lambda})^{2}}{\lambda}
\geq C_{1}\inf_{i\geq 1}a_{i}\,\dfrac{\left(\sqrt{\overline{z}+\delta}-\sqrt{\overline{z}+
\frac{\delta}{2}} \right)^{2}}{\overline{z}+\frac{\delta}{2}},
$$ completing the proof.
\end{proof}

Next, we present a new lower bound estimation for the energy dissipation in the case where $c_1$ is 'too small'.
\begin{lem}\label{lem:D for small c_1 general sum a_i c_i}
Let $\br{a_i}_{i\in\N},\br{Q_i}_{i\in\N}$ be the coagulation and detailed balance coefficients for the Becker-D\"oring equations. Assume that
$$\overline{Q}=\sup_{i}\frac{Q_{i}}{Q_{i+1}}<+\infty \quad \underline{Q}=\inf_{i}\frac{Q_{i}}{Q_{i+1}}<+\infty$$
$$\overline{a}=\sup_{i}\frac{a_{i}}{a_{i+1}}<+\infty \quad \underline{a}=\inf_{i}\frac{a_{i}}{a_{i+1}}<+\infty,$$
and let $\bm{c}$ be a non-negative sequence such that
$$c_1 < \delta$$
for some $\delta>0$. Then,
$$ {\overline{D}(\bm{c})} \geq \underline{Q}\underline{a} \pa{\sum_{i=1}^\infty a_ic_i-a_1\delta}-2\sqrt{\delta} \sqrt{\overline{Q}\overline{a}}\pa{\sum_{i=1}^\infty a_ic_i}.$$
\end{lem}
\begin{proof} Expanding the square, one has 
$$\overline{D}(\bm{c}) =c_1\sum_{i=1}^\infty a_i c_i + \sum_{i=1}^\infty a_i \frac{Q_i}{Q_{i+1}}c_{i+1} - 2\sqrt{c_1}\sum_{i=1}^\infty a_i\sqrt{\frac{Q_i}{Q_{i+1}}}\sqrt{c_ic_{i+1}}$$
so that
\begin{multline*}
\overline{D}(\bm{c}) \geq \underline{Q}\underline{a} \pa{\sum_{i=2}^\infty a_ic_i}-2\sqrt{c_1} \sqrt{\overline{Q}\overline{a}}\sqrt{\sum_{i=2}^\infty a_ic_i}\sqrt{\sum_{i=1}^\infty a_ic_i}\\
\geq \underline{Q}\underline{a} \pa{\sum_{i=1}^\infty a_ic_i-a_1\delta}-2\sqrt{\delta} \sqrt{\overline{Q}\overline{a}}\pa{\sum_{i=1}^\infty a_ic_i},\end{multline*}
which is the desired result.
\end{proof}
\begin{prp}\label{prp:uniform lower bound on D if c_1 is small}
Let $\br{a_i}_{i\in \N},\br{Q_i}_{i\in\N}$ be the coagulation and detailed balance coefficients for the Becker-D\"oring equations. Assume that
$$\overline{Q}=\sup_{i}\frac{Q_{i}}{Q_{i+1}}<+\infty \quad \underline{Q}=\inf_{i}\frac{Q_{i}}{Q_{i+1}}<+\infty.$$
Let $\bm{c}$ be a non-negative sequence with finite total density $\varrho$. Then:
\begin{enumerate}[(i)]
\item If $a_i=i$ then there exists a $\delta_1>0$, depending only on $\overline{Q},\underline{Q}$ and $\varrho$ such that if $c_1 < \delta_1$ then 
$$ {\overline{D}(\bm{c})} \geq \frac{\underline{Q}\varrho}{4}.$$
\item If $a_i=i^\gamma$ for $\gamma<1$ and there exists $\beta>1$ such that $M_\beta< +\infty$, then there exists $\delta_1>0$, depending only on $\overline{Q},\underline{Q},\varrho$ and $M_\beta$  such that if $c_1 < \delta_1$ then 
$$ {\overline{D}(\bm{c})} \geq \frac{\underline{Q}\varrho^{\frac{\beta-\gamma}{\beta-1}}}{4M_\beta^{\frac{1-\gamma}{\beta-1}}}.$$
\end{enumerate}
\end{prp}
\begin{proof}
Both $(i)$ and $(ii)$ will follow immediately from Lemma \ref{lem:D for small c_1 general sum a_i c_i} and a suitable choice of $\delta_1$ .
Indeed, for $(i)$ we notice that 
$$\underline{Q}\underline{a} \pa{\sum_{i=1}^\infty a_ic_i-a_1\delta}-2\sqrt{\delta} \sqrt{\overline{Q}\overline{a}}\pa{\sum_{i=1}^\infty a_ic_i}=\frac{\underline{Q}}{2}\pa{\varrho-\delta}-2\sqrt{\delta}\sqrt{\overline{Q}}\varrho,$$
where we have used the notations of Lemma \ref{lem:D for small c_1 general sum a_i c_i}. As the above is less than $\frac{\underline{Q}\varrho}{2}$ and converges to it as $\delta$ goes to zero, we can find $\delta_1$ that satisfies the desired result.

For $(ii)$ we notice that the following interpolation estimate
$$\varrho = \sum_{i=1}^\infty ic_i \leq \pa{\sum_{i=1}^\infty i^\gamma c_i}^{\frac{\beta-1}{\beta-\gamma}}\pa{M_\beta}^{\frac{1-\gamma}{\beta-\gamma}}$$
along with the fact that $\sum_{i=1}^\infty i^\gamma c_i \leq \varrho$ implies that
$$\underline{Q}\underline{a} \pa{\sum_{i=1}^\infty a_ic_i-a_1\delta}-2\sqrt{\delta} \sqrt{\overline{Q}\overline{a}}\pa{\sum_{i=1}^\infty a_ic_i}
\geq \frac{\underline{Q}}{2}\pa{\frac{\varrho^{\frac{\beta-\gamma}{\beta-1}}}{M_\beta^{\frac{1-\gamma}{\beta-1}}}-\delta} -2\sqrt{\delta}\sqrt{\overline{Q}}\varrho,$$
from which the result follows.
\end{proof}
We are finally ready to complete the proof of Theorem \ref{thm:main thm fe}:
\begin{proof}[Proof of part \eqref{enm: uniform lower bound} of Theorem \ref{thm:main thm fe}]
This follows immediately form Propositions \ref{prp:uniform lower bound on D if c_1 is large} and \ref{prp:uniform lower bound on D if c_1 is small}.
\end{proof}

Now that we have our general functional inequality at hand one may wonder how sharp is this method of using the log-Sobolev inequality? Perhaps we were too coarse in our estimation, and Cercignani's conjecture \emph{is} valid in the case $a_i=i^\gamma$ with $\gamma<1$ under the restrictions of Theorem \ref{thm:main thm fe}. The answer, surprisingly, is that this method is optimal, as we shall see in the next subsection.

 \subsection{Optimality of the Results}

This subsection is devoted to showing that unlike the case $a_i=i$, the case $a_i=i^\gamma$ when $\gamma<1$ \emph{admits no Cercignani's Conjecture}, even if $c_1$ is bounded appropriately. This is stated in Theorem \ref{thm:optimal}.
\begin{proof}[Proof of Theorem \ref{thm:optimal}] 
We start by choosing $a_i=i^\gamma$, $\gamma<1$, and $Q_i=e^{-\lambda(i-1)}$ $(i \geq 1)$ for some $\lambda \geq 0$. We will show the desired result by constructing a family of {non-negative} sequences, $\br{\bm{c}^{(\varepsilon)}}_{\varepsilon>0}$ with a fixed mass $\varrho$ such that
$$\lim_{\varepsilon\rightarrow 0 }\frac{D\pa{\bm{c}^{(\varepsilon)}}}{H\pa{\bm{c}^{(\varepsilon)}|\Q}}=0.$$
 Let $\xi>0$ be such that
$${\frac{\varrho}{2}}  = \sum_{i=1}^\infty i e^{\lambda}e^{-\xi i} = \frac{e^{\lambda-\xi}}{\pa{1-e^{-\xi}}^2}. $$
Consider the sequence {$\bm{c}^{(\varepsilon)}=\br{c_{i}^{(\varepsilon)}}$ given by}
$$c^{(\varepsilon)}_i=e^{\lambda}e^{-\xi i}+A_\varepsilon e^{-\varepsilon i}, \qquad i \in \N$$
where $0<\varepsilon$ is small and $A_\varepsilon$ is chosen such that the mass of the sequence $\bm{c}^{(\varepsilon)}$ is $\varrho$, i.e. 
$A_\varepsilon =\frac{\varrho}{2}e^{\varepsilon}\pa{1-e^{-\varepsilon}}^2.$
Next, as $\frac{Q_i}{Q_{i+1}}=e^\lambda$  {for any $i \geq 1$}, we see that
$$\frac{Q_i}{Q_{i+1}}c^{(\varepsilon)}_{i+1}-c^{(\varepsilon)}_1c^{(\varepsilon)}_i = e^{2\lambda} e^{-\xi(i+1)}+A_\varepsilon e^{\lambda} e^{-\varepsilon(i+1)} - e^{2\lambda}e^{-\xi (i+1)} - A_\varepsilon e^{\lambda}\pa{e^{-\xi i -\varepsilon}+e^{-\varepsilon i -\xi}} 
-A_\varepsilon^2 e^{-\varepsilon(i+1)}$$
$$=A_\varepsilon e^{\lambda}e^{-\varepsilon (i+1)} \pa{1- e^{-(\xi-\varepsilon)}  -e^{-(\xi-\varepsilon)i}-A_\varepsilon e^{-\lambda}}  >0$$
for $\varepsilon$ small enough  {depending only on $\lambda,\xi$ and $\varrho$ but not on $i$.} Additionally, one can easily verify that
$$\frac{Q_i c^{(\varepsilon)}_{i+1}}{Q_{i+1}c^{(\varepsilon)}_1c^{(\varepsilon)}_i}\leq e^{\lambda}\pa{ 1+ \frac{1}{A_\varepsilon}}.$$
As such, setting {$B_{z,\gamma}=\sum_{i=1}^\infty  i^\gamma e^{-z i} $ for any $z > 0$}, we find that
\begin{equation}\label{eq:no cerc conject for gamma<1 I}
\begin{gathered}
D\pa{\bm{c}^{(\varepsilon)}} =\sum_{i=1}^\infty i^\gamma \pa{\frac{Q_i}{Q_{i+1}}c^{(\varepsilon)}_{i+1}-c^{(\varepsilon)}_i}\log\pa{\frac{Q_i c^{(\varepsilon)}_{i+1}}{Q_{i+1}c^{(\varepsilon)}_1c^{(\varepsilon)}_i}} \\
\leq A_\varepsilon e^{\lambda}B_{\varepsilon,\gamma} \log\pa{e^{\lambda}\pa{1+\frac{1}{A_\varepsilon}}}\pa{\pa{1-A_\varepsilon e^{-\lambda}}e^{-\varepsilon}-e^{-\xi}} \\
 - A_\varepsilon e^{\lambda} {B_{\xi,\gamma}}\log\pa{e^{\lambda-\varepsilon}\pa{1+\frac{1}{A_\varepsilon}}}.
\end{gathered}
\end{equation}
As $A_{\varepsilon}\approx  {\frac{\varrho}{2}}\varepsilon^2$ when $\varepsilon$ approaches zero, and $B_{\varepsilon,\gamma}$ is of order $\varepsilon^{-(1+\gamma)}$ (see Lemma \ref{lemapp:order of B_epsilon_gamma} in Appendix \ref{app:additional useful}) we conclude that 
$$\lim_{\varepsilon\rightarrow 0}D\pa{\bm{c}^{(\varepsilon)}}=0.$$
Lastly, we turn our attention to the relative  free energy. We start by denoting by $\overline{\xi}>0$ the unique parameter for which 
$$\varrho ={e^{\lambda}} \sum_{i=1}^\infty i e^{-\overline{\xi} i}.$$
Clearly, $\overline{\xi}<\xi$ {and the associated equilibrium with mass $\varrho$ is $\Q_{i}=e^{\lambda}e^{-\overline{\xi}i}$.}  {Since, for any fixed $i \geq 1$, it holds
$$\lim_{\varepsilon \to 0}c_{i}^{(\varepsilon)}=c_{i}^{(0)}=e^{\lambda}e^{-\xi i}$$
}using Fatou's lemma we can conclude that
$$\liminf_{\varepsilon\rightarrow 0} H\pa{c^{(\varepsilon)}|\Q} \geq H\pa{{\bm{c}^{(0)}}| \Q}>0$$
as $\bm{c}^{(0)}\not= \Q$.
\end{proof}
\begin{rem}\label{rem:q_s infinite}
We Notice the following:
\begin{itemize}
\item In the example we provided $z_s=e^{\lambda}<+\infty$ but $\varrho_s=+\infty$. This, however, is not a great obstacle as all our proofs rely on some \emph{positive} distance form $z_s$ and $\varrho_s$, and can be reformulated accordingly. 
\item The constructed sequence $\bm{c}^{(\varepsilon)}$ satisfies
$$\sup_{\varepsilon}\sum_{i=1}^{\infty}i^{\beta}\,c_{i}^{(\varepsilon)}=+\infty$$
for any $\beta > 1$. Thus, the conclusion of part \eqref{enm: gamma<1 almost cerc conjecture} of Theorem \ref{thm:main thm fe} does not apply to it. Actually, one can easily check that $\lim_{\varepsilon\to 0}\frac{D(\bm{c}^{(\varepsilon)})}{\left(H(\bm{c}^{(\varepsilon)}|\Q)\right)^{s}}=0$
for any $s > 0.$
\end{itemize}
\end{rem}

\subsection{Inequalities with Exponential Moments}
\label{sec:exp moment extra}

Up to now, we have avoided using exponential moments in any of our functional inequalities. In this section we will show that when $0\leq \gamma<1$, under the additional assumption of a bounded exponential moment, one can obtain a far better functional inequality between $\overline{D}(\bm{c})$ and $H(\bm{c}|\Q)$, extending the result given by Jabin and Neithammer.\\
The key idea in this section is to avoid using the interpolation
inequality \eqref{eq:interpolation for gamma<1} and replace it with
one that involved an exponential weight.
\begin{prp}\label{prop:exp interpolation}
Let $\bm{f}$ be a non-negative sequence and let $0\leq \gamma<1$. Assume that there exists $\mu \in (0,4\log 2)$ such that
$$\sum_{i=1}^\infty e^{\mu i}f_i={M_{\mu}^{\exp}(\bm{f})} <+\infty.$$
Then,
\begin{equation}\label{eq:exp interpolation}
M_\gamma {(\bm{f})} \geq 
\frac{M_1 {(\bm{f})}}{2\pa{\frac{2}{\mu}\log\pa{\frac{4 {M_{\mu}^{\exp}(\bm{f})}}{\mu e M_1(\bm{f})}}}^{1-\gamma}} \end{equation}
where $M_\alpha {(\bm{f})}$ denotes the $\alpha-$moment of $\bm{f}$. 
\end{prp}
\begin{proof}
For simplicity, we will use the notation of $M_1$ and $M^{\exp}_{\mu}$ instead of $M_1(\bm{f})$ and $M_{\mu}^{\exp}(\bm{f})$. We start with the simple inequality
\begin{equation}\label{eq:M1 interpolation}
\begin{split}
M_1&=\sum_{i=1}^\infty i f_i = \sum_{i=1}^N i^{1-\gamma}i^\gamma f_i + \sum_{i=N+1}^\infty ie^{-\frac{\mu i}{2}}e^{-\frac{\mu i}{2}}e^{\mu i}f_i\\
&\leq N^{1-\gamma}M_\gamma + \frac{2e^{-\frac{\mu(N+1)}{2}}}{\mu e} {M_{\mu}^{\exp}}, \qquad \forall N \in \N\end{split}\end{equation}
where we used the fact that $\sup_{x\geq 0} xe^{-\lambda x}=\frac{1}{\lambda e}$ {for any $\lambda >0$}. Our goal will be to choose a particular $N$ to plug in the inequality above to conclude the desired result. Again, using the supremum of $g(x)=xe^{-\lambda x}$, we conclude that
 $$M_{1} \leq \frac{1}{\mu\,e}M_{\mu}^{\exp}.$$ 
As $\mu < 4\log 2$ we find that
$$M_1 < \frac{4{M_{\mu}^{\exp}}}{\mu e^{1+\frac{\mu}{2}}}.$$ 
from which we conclude that $N=\left[\frac{2}{\mu}\log\pa{\frac{4 {M_{\mu}^{\exp}}}{\mu e M_1}}\right] \geq 1$. Plugging this $N$ into \eqref{eq:M1 interpolation} we see that ${e^{-\frac{\mu (N+1)}{2}} \leq  \frac{\mu e M_1}{4{M_{\mu}^{\exp}}}, }$
and as such
$$M_{\gamma} \geq N^{\gamma-1}\frac{M_{1}}{2}$$
and the result follows. \end{proof}

With this proposition at hand, we are prepared to show part \eqref{enm:exp func ineq} of Theorem \ref{thm:rate of convergence exp}.
\begin{proof}[Proof of part \eqref{enm:exp func ineq} of Theorem \ref{thm:rate of convergence exp}]
Without loss of generality we may assume that $\mu\in(0,4\log 2)$. Introduce the sequence $\bm{f}=\br{f_i}$ where
$$f_{i}=Q_{i}\left(\sqrt{\dfrac{c_{1}c_{i}}{Q_{i}}}-\sqrt{\dfrac{c_{i+1}}{Q_{i+1}}}\right)^{2}, \qquad i \geq 1.$$ 
Following the same proof as presented in Lemma \ref{lem:D-upper-bound} we find that
$$M_{\mu}^{\exp}(\bm{f}) \leq 2\pa{c_1+z_s\sup_j \frac{\alpha_j}{\alpha_{j+1}}}M_{\mu}^{\exp}(\bm{c}).$$
Thus, using the simple fact that $M_{\alpha}(\bm{f})=\overline{D}_\alpha(\bm{c})$, for any $\alpha>0$, together with Proposition \ref{prop:exp interpolation} and parts \eqref{enm: gamma=1 cerc conjecture} and \eqref{enm: uniform lower bound} of Theorem \ref{thm:main thm fe} yield the desired functional inequality.
\end{proof}

\section{Rate of Convergence to Equilibrium}
\label{sec: energy dissipation for BD}

In this section we will use all the information we gathered so far to
show the proof of Theorems
\ref{thm:rate of convergence} and part \eqref{enm:exp mom rate} of Theorem \ref{thm:rate of convergence exp}, giving an explicit rate of convergence
to equilibrium for the Becker-D\"oring equations.

The convergence is an immediate consequence of Theorem \ref{thm:main thm fe} and part \eqref{enm:exp func ineq} of Theorem \ref{thm:rate of convergence exp}, yet we provide a proof here for the sake of completion and to show that we can find all the constants explicitly.
\begin{proof}[Proof of Theorem \ref{thm:rate of convergence}]
Due to Theorem \ref{thm:main thm fe} we conclude the following differential inequality:
\begin{equation}\label{eq:convergence rate proof 0}
\frac{\d}{\d t}H(\bm{c}(t)|\Q) \leq \begin{cases}
-\min\pa{KH(\bm{c}(t)|\Q),\varepsilon} & \gamma=1. \\
-\min\pa{KH(\bm{c}(t)|\Q)^{\frac{\beta-\gamma}{\beta-1}},\varepsilon} & 0\leq \gamma <1,
\end{cases}
\end{equation}
for appropriate $K$ and $\varepsilon$. We claim that there exists $t_0 \geq 0$ such that for all $t\geq t_0$
\begin{equation}\label{eq:convergence rate proof I}
H(c(t)|\Q) \leq \begin{cases}
\frac{\varepsilon}{K} & \gamma=1 \\
\pa{\frac{\varepsilon}{K}}^{\frac{\beta-1}{\beta-\gamma}} & 0\leq \gamma <1. 
\end{cases}
\end{equation}
Indeed, if $H(\bm{c}(t))|\Q)$ is bigger then the appropriate constants in $[0,t]$ then
$$\frac{\d}{\d s}H(\bm{c}(s)|\Q) \leq - \varepsilon \quad \forall s\in (0,t) ,$$
implying that
\begin{equation*}
H(\bm{c}(t)|\Q) \leq H(\bm{c}(0)|\Q) -\varepsilon t.
\end{equation*}
We define 
$$t_0=\begin{cases}
\min\pa{0,\frac{H(\bm{c}(0)|\Q)-\frac{\varepsilon}{K}}{\varepsilon}} & \gamma=1\\
\min\pa{0,\frac{H(\bm{c}(0)|\Q)-\pa{\frac{\varepsilon}{K}}^{\frac{\beta-1}{\beta-\gamma}}}{\varepsilon}} & 0\leq \gamma<1.
\end{cases}$$
and find that $H(\bm{c}(t_0)|\Q)$ satisfies the appropriate inequality in \eqref{eq:convergence rate proof I}. As $H(\bm{c}(t)|\Q)$ is decreasing, we conclude that \eqref{eq:convergence rate proof I} is valid for any $t\geq t_0$.

With this in hand, along with \eqref{eq:convergence rate proof 0}, we have that for all $t\geq t_0$:
$$H(\bm{c}(t)|\Q) \leq \begin{cases}
H(\bm{c}(t_0)|\Q)e^{-K(t-t_0)} & \gamma=1 \\
\frac{1}{\pa{H(\bm{c}(t_0)|\Q)^{\frac{\gamma-1}{\beta-1}}+\frac{1-\gamma}{\beta-1}K(t-t_0)}^{\frac{\beta-1}{1-\gamma}}} & 0\leq \gamma <1.
\end{cases}$$
As 
$$H(\bm{c}(t_0)|\Q)=\begin{cases}
\min\pa{H(\bm{c}(0)|\Q),\frac{\varepsilon}{K}} & \gamma=1 \\
\min\pa{H(\bm{c}(0)|\Q),\pa{\frac{\varepsilon}{K}}^{\frac{\beta-1}{\beta-\gamma}}} & 0\leq \gamma <1,
\end{cases}$$
and $t_0$ is given explicitly we conclude that
$$ {C\left(H(\bm{c}(0)|\Q)\right)}=\begin{cases}
H(\bm{c}(0)|\Q)& \gamma=1, t_0=0 \\
\frac{\varepsilon}{K}e^{K\frac{H(\bm{c}(0)|\Q)-\frac{\varepsilon}{K}}{\varepsilon}}& \gamma=1, t_0>0 \\
H(\bm{c}(0)|\Q) & 0\leq \gamma <1, t_0=0\\
\pa{\frac{\varepsilon}{K}}^{\frac{\gamma-1}{\beta-\gamma}}-\frac{1-\gamma}{\beta-1}K\frac{H(\bm{c}(0)|\Q)-\pa{\frac{\varepsilon}{K}}^{\frac{\beta-1}{\beta-\gamma}}}{\varepsilon} & 0\leq \gamma <1, t_0>0,
\end{cases}$$
completing the proof.
\end{proof}
\begin{proof}[Proof of part \eqref{enm:exp mom rate} of Theorem \ref{thm:rate of convergence exp}]
This follows form part \eqref{enm:exp func ineq} of Theorem \ref{thm:rate of convergence exp} by the same methods used in the above proof and the fact that
$$\sup_{t \geq 0}{M}_{\mu^\prime}^{\exp}(\bm{c}(t))<+\infty$$
for some $0<\mu^\prime<\mu$ (a known result from \cite{JN03}). 
\end{proof}

\section{Consequences for General Coagulation and Fragmentation
  Models}\label{sec:application to general CF}

The Becker-D\"oring equations \eqref{eq:BD} are derived under the
assumption that the only relevant reactions taking place are those
between monomers and clusters of any size. One can obtain a more
general model by taking into account reactions between clusters of any
size. Keeping the notation of the introduction, this means that we
consider reactions of the type
\begin{equation*}
  \{i\} + \{j\} \rightleftharpoons \{i+j\}
\end{equation*}
for any positive integer sizes $i$ and $j$. We assume their
coagulation rate (i.e., the reaction from left to right) is determined
by a coefficient we call $a_{i,j}$, and their fragmentation rate (the
reaction from right to left) by a coefficient called $b_{i,j}$. These
coefficients are always assumed to be nonnegative (as before) and
symmetric in $i,j$ (that is, $a_{i,j} = a_{j,i}$ and
$b_{i,j} = b_{j,i}$ for all $i$, $j$). The corresponding to
eq. \eqref{eq:BD} is then
\begin{equation}
  \label{eq:CF}
  \frac{\d}{\d t} c_i(t) = \frac12 \sum_{j=1}^{i-1} W_{j, i-j}(t)
                  - \sum_{j=1}^\infty W_{i,j}(t), \qquad i\in \N.
\end{equation}
where
\begin{equation}
  \label{eq:def-Wij}
  W_{i,j}(t) := a_{i,j}\, c_i(t) c_j(t) - b_{i,j}\, c_{i+j}(t)
  \qquad i \in \N.
\end{equation}
The system \eqref{eq:BD} is then a particular case of \eqref{eq:CF}
obtained by choosing $a_{i,j}$, $b_{i,j}$ as
\begin{gather}
  a_{i,j} = b_{i,j} = 0 \quad \text{ when $\min\{i,j\} \geq 2$},
  \\
  \label{eq:ai-aii}
  a_{1,1} := 2 a_1,
  \qquad
  a_{i,1} = a_{1,i} = a_i \quad \text{ for $i \geq 2$},
  \\
  \label{eq:bi-bii}
  b_{1,1} := 2 b_2,
  \qquad
  b_{i,1} = b_{1,i} = b_{i+1} \quad \text{ for $i \geq 2$}.
\end{gather}
The mathematical theory of this full system is much less complete than
that of \eqref{eq:BD}. Well-posedness of mass-conserving solutions has
been studied  {in} \cite{BC90}, and there are a number
of works on asymptotic behaviour, for instance
\cite{CC92,CC94,C05,C07}, but it is still not fully understood. To
start with, it is unclear whether equilibria of \eqref{eq:CF} are
unique or not (when they exist). A common physical condition imposed
on the coefficients $a_{i,j}$, $b_{i,j}$ which avoids this problem is
that of \emph{detailed balance}: we say it holds when there exists a
sequence $\{Q_i\}_{i \geq 1}$ of strictly positive numbers such that
\begin{equation}
  \label{eq:CFdb}
  a_{i,j} Q_i Q_j = b_{i,j} Q_{i+j} \quad \text{ for any } i,j,
\end{equation}
where we always further assume without loss of generality that
$Q_1 = 1$.  This is the analogue of \eqref{eq:Qi}, but in this case it
needs to be imposed as a condition since numbers $Q_i$ satisfying
\eqref{eq:CFdb} cannot always be found (unlike in the Becker-D\"oring
case). If we assume \eqref{eq:CFdb} then equilibria \eqref{eq:CF}
exist and have the same form \eqref{eq:equilibria} as in the
Becker-D\"oring case, and a similar phase transition in the long-time
behaviour has been rigorously proved in some cases (see
\cite{CC92,CC94,C05,C07} for more details). However, even with
detailed balance the long-time behaviour is in general not understood
except in particular cases. If clusters larger than a given size $N$
do not react among themselves (that is, if $a_{i,j} = b_{i,j} = 0$
whenever $\min\{i,j\} > N$) the system is known as the
\emph{generalised Becker-D\"oring system}, and has been studied in
\cite{Costa98,C05}. For coefficients $a_{i,j}$ given by
\begin{equation}
  \label{eq:aij}
  a_{i,j} = i^\gamma j^\eta + i^\eta j^\gamma \quad \text{ for any } i,j,
\end{equation}
with $\eta \leq 0 \leq \gamma$ and $\gamma + \eta \leq 1$, the
asymptotic behaviour was identified in \cite{C07} and a constructive
(though probably far from optimal) rate of convergence to equilibrium
was given. Very little is known about the asymptotic behaviour for
coefficients of the type \eqref{eq:aij} with $\gamma, \eta > 0$ and
$\gamma + \eta \leq 1$. In this case the size of $a_{i,i}$ is larger
than that of $a_{i,1}$ and the system \eqref{eq:CF} may be behave
quite differently from \eqref{eq:BD}.

The purpose of this section is to clarify whether any of the
functional inequalities investigated in this paper can shed new light
on the behaviour of solutions to \eqref{eq:CF}. Assuming the detailed
balance condition \eqref{eq:CFdb}, along a solution
$\bm{c}(t) = \{c_i(t)\}_{i \geq 1}$ to \eqref{eq:CF} we have
\begin{multline}
  \label{eq:H-thm-CF}
  \frac{\d}{\dt} H(\bm{c}(t))
  = - D_{\mathrm{CF}}(\bm{c}(t))
  \\
  := -\frac12 \sum_{i,j=1}^\infty a_{i,j} Q_i Q_j
  \left(\frac{c_i c_j}{Q_i Q_j} - \frac{c_{i+j}}{Q_{i+j}} \right)
  \left(\log \frac{c_i c_j}{Q_iQ_j}
    - \log \frac{c_{i+j}}{Q_{i+j}} \right)
  \\
  \leq
  - \sum_{i=1}^\infty a_{i} Q_i
  \left(\frac{c_i c_1}{Q_i} - \frac{c_{i+1}}{Q_{i+1}} \right)
  \left(\log \frac{c_i c_1}{Q_i}
    - \log \frac{c_{i+1}}{Q_{i+1}} \right)
  = D(\bm{c}(t)) \leq 0
\end{multline}
(see \cite{C07} for a rigorous proof) where $a_{i}$ are defined by
\eqref{eq:ai-aii} for any $i \geq 1$. Hence the free energy is also a
Lyapunov functional for \eqref{eq:CF}, and it dissipates at a
\emph{faster} rate than for the Becker-D\"oring equations (since more
types of reactions are allowed). As such, it is reasonable to think
that the inequalities from Section \ref{sec:EDI} can be useful also in
this case. This turns out to be true, and some improvements can be
made on existing results. However, it also turns out that our results
are not able to extend the range of possible coefficients for which
convergence to a particular subcritical equilibrium can be proved; we
cannot give any new results for coefficients such as \eqref{eq:aij}
with $\gamma, \eta > 0$ and $\gamma + \eta \leq 1$.

One of the main obstacles in applying our results to equation
\eqref{eq:CF} is that, unlike for the Becker-D\"oring equations, the
moments of solutions to the general coagulation and fragmentation
system are not known to be bounded. One can for example say the
following about integer moments (this result can easily be extended to
non-integer powers by interpolation, and was known from the early
works in the topic \cite{CC92,CC94}). From this point onward we will
assume that
\begin{equation}
  \label{eq:aij-power}
  a_{i,j}=i^\gamma j^\eta+i^\eta j^\gamma
  \quad \text{ for $i,j \in \N$},
\end{equation}
with $\eta \leq \gamma$ and $0 \leq \lambda := \gamma+\eta \leq 1$.

\begin{lem}\label{lem:moments bounds in general CF}
  Let $k\in\N$ and let $\bm{c} = \bm{c}(t)=\br{c_{i}(t)}_{i \in \N}$
  be a solution with mass $\varrho$ to the coagulation and
  fragmentation system \eqref{eq:CF} with coefficients
  satisfying \eqref{eq:aij-power}. Then
  \begin{equation}\label{eq:moments bounds in general CF}
    M_k(\bm{c}(t))
    \leq
    \begin{cases}
      \pa{M_k({\bm{c}(0)})+\frac{1-\lambda}{k-1}
        \pa{2^k-2}\varrho^{\frac{1-\gamma}{k-1}}t}^{\frac{k-1}{1-\lambda}}
      & \text{ if } \quad 0<\lambda<1 \\
      M_k({\bm{c}(0)}) \exp\left(2\pa{2^k-2}\varrho t\right)
      & \text{ if } \quad \lambda=1
    \end{cases}
  \end{equation}
  where $M_{p}(\bm{c}(t)) := \sum_{i=1}^{\infty}i^{p}c_{i}(t)$ for any $p \geq 0$, $t \geq 0$.
\end{lem}

\begin{proof}
  We give a formal proof for completeness; a rigorous one can be
  obtained by standard approximation methods, and can be found in
  \cite{BC90}. To simplify the notation and since $\bm{c}(t)$ is
  fixed, we denote $M_{j}(t)=M_{j}(\bm{c}(t))$ for any $j\geq 1,$
  $t \geq 0$. One can check the following weak formula for the
  integral of the right hand side of \eqref{eq:CF} against a test
  sequence $\br{\phi(i)}_{i}$:
  $$\sum_{i=1}^\infty
  \phi(i)\pa{\frac{1}{2}\sum_{j=1}^{i-1}W_{i-j,j}-\sum_{j=1}^\infty
    W_{i,j}} = \frac{1}{2}\sum_{i,j}^\infty
  \pa{\phi(i+j)-\phi(i)-\phi(j)}W_{i,j}.$$
  Applying this to $\phi(i) := i^k$, neglecting the negative
  contribution of the fragmentation terms and using the binomial
  formula one obtains
  $$\frac{\d}{\d t}M_k(t) \leq \sum_{l=1}^{k-1}
  {k \choose l}
  M_{l+\gamma}(t)M_{k-l+\eta}(t) \qquad \forall t \geq 0.$$
  Next, we use the interpolation
  $$M_{\delta}{(t)} \leq M_1^{\frac{k-\delta}{k-1}}{(t)}
  M_k^{\frac{\delta-1}{k-1}}{(t)}$$
  where $1<\delta<k$, to find that
  $$M_{l+\gamma}(t)M_{k-l+\eta}(t) \leq M_1(t)^{\frac{k-\lambda}{k-1}}M_{k}(t)^{\frac{k+\lambda-2}{k-1}}.$$
  Thus,
  $$\frac{\d}{\d t}M_k(t) \leq \pa{2^k-2}\varrho^{\frac{k-\lambda}{k-1}}M_{k}^{\frac{k+\lambda-2}{k-1}}(t) \qquad \forall t\geq 0$$
  and the result follows from this differential inequality.
\end{proof}
With the above at hand, we are now able to use the theory developed in
the previous sections for the Becker-D\"oring equations in order to
conclude a rate of convergence to equilibrium in the general setting
of coagulation and fragmentation equations. Our main theorem is the
following:

\begin{thm}[Asymptotic behaviour of the coagulation-fragmentation system]
  \label{thm:almost cerc for CF}
  Let ${\br{a_{i,j}}_{i,j\in\N}, \br{b_{i,j}}_{i,j\in\N}}$ be the
  coagulation and fragmentation coefficients for equation
  \eqref{eq:CF}, and assume that the detailed balance condition
  \eqref{eq:CFdb} holds. Assume that
  \begin{equation}
    a_{i,j}=i^\gamma + j^\gamma,\label{eq:aij2}
  \end{equation}
  for some $0\leq\gamma<1$ and that
  ${\br{Q_i}_{i\in\N}}$ satisfies Hypothesis
  \ref{hyp: Qi-form}. Assume in addition that
  $M_k(\bm{c}(0))<+\infty$ for some $k\in\N$,
  $k>1$. Then
  \begin{equation}\label{eq:rate of convergence general CF}
    H(\bm{c}(t)|\Q) \leq \frac{1}{\pa{C_1+C_2\log t}^{\frac{k-1}{1-\gamma}}}
  \end{equation}
  where $C_1,C_2>0$ are constants depending only on
  $H(\bm{c}(0)|\Q), z_s, \varrho, \br{\alpha_i}_{i\in\N}, k,
  \gamma$ and $M_k(\bm{c}(0))$.
\end{thm}

\begin{proof}
  Assume for the moment that $a_{i,j}$ is of the form \eqref{eq:aij},
  in order to see why the proof only works for coefficients of the
  form \eqref{eq:aij2}.

  Fix $\delta > 0$ such that
  $0 < \delta < \overline{z} < \zs - \delta$. We use the observation
  \eqref{eq:H-thm-CF} that
  $D_{\mathrm{CF}}(\bm{c}(t)) \geq D(\bm{c}(t))$ at all times
  $t \geq 0$ (defining $\{a_i\}_{i \in \N}$ by
  \eqref{eq:ai-aii}). Using Theorem \ref{thm:main thm fe} (actually,
  its more detailed forms in equation \eqref{eq:almost-cercignani} and
  Proposition \ref{prp:uniform lower bound on D if c_1 is small}) we
  obtain the following:
  \begin{align*}
    \ddt H(\bm{c}(t) | \Q)
    &= -D_{\mathrm{CF}}(\bm{c}(t))
    \leq -D(\bm{c}(t))
    \\
    &\leq
    \begin{cases}
      - C M_k(\bm{c}(t))^{\frac{\gamma-1}{k-1}}
      H({\bm{c}}(t)|\Q)^{\frac{k- \gamma}{k-1}}
      & \quad \text{ if } \delta < c_1(t) < \zs-\delta
      \\
      - C M_k(\bm{c}(t))^{\frac{\gamma-1}{k-1}}  &  \text{ if $c_1(t) < \delta$ or $c_1(t) \geq \zs-\delta$.}
    \end{cases}
    \\
    &\leq
    -C_0 M_k(\bm{c}(t))^{\frac{\gamma-1}{k-1}}
      H({\bm{c}}(t)|\Q)^{\frac{k- \gamma}{k-1}}
    \end{align*}
    for some constant $C_0 > 0$ that depends also on
    $H(\bm{c}(0)|\Q)$. Using Lemma \ref{lem:moments bounds in general
      CF} this implies
    \begin{equation*}
      \ddt H(\bm{c}(t) | \Q)
      \leq
      - \frac{C_0}{
        \left(
          M_k(\bm{c}(0)) + \frac{1-\lambda}{k-1}(2^k - 2)
          \varrho^{\frac{k-\lambda}{k-1}} t
        \right)^{\frac{1-\gamma}{1-\lambda}}
      }
      H({\bm{c}}(t)|\Q)^{\frac{k- \gamma}{k-1}} \qquad t \geq 0.
    \end{equation*}
    This implies decay of $H(\bm{c}(t))$ only when $\lambda = \gamma$,
    that is, when $\eta=0$ (since $\lambda = \gamma + \eta$). Solving
    the differential inequality yields the result.
  \end{proof}
  
\begin{rem}
  The same decay rate was obtained in \cite{C07} by means of the
  particular case of inequality \eqref{eq:almost cerc conjecture fe}
  for $k = 2-\gamma$. Here we obtain slightly different decay rates by
  assuming higher moments of the initial data $\bm{c}(0)$ are finite,
  but the method does not seem to give a better decay than a power of
  $\log t$ in any case.
\end{rem}

\section{Final Remarks.}\label{sec:final remarks}

In this final section we gather a few remarks and open problems associated to this paper:

\textit{\textbf{The assumption that $c_1(t)$ is in the 'good' region}} given by \eqref{eq:good c_1} is not so far-fetched. Indeed, as long as we know that the energy decreases in some given way, we can apply the following Csisz\'ar-Kullback inequality, \eqref{eq:csiszar kullback ineq},
$$\sum_{i=1}^\infty \abs{c_i(t)-\Q_i }\leq \sqrt{2\varrho H({\bm{c}}(t)|\Q)}$$
and obtain that if $H(\bm{c}(t_0)|\Q)$ is small enough then for any $t>t_0$
$$\overline{z}-H(\bm{c}(t_0)|\Q) \leq c_1(t) \leq \overline{z}+H(\bm{c}(t_0)|\Q).$$

\textit{\textbf{The hypothesis on the form of $\br{Q_i}_{i\in\N}$}}, stated in Hypothesis \ref{hyp: Qi-form}, was used explicitly in our work in order to obtain very \emph{quantitative} estimations on the constants appearing in our theorems. As one can see from the proofs, this hypothesis can be relaxed - but the price one must pay is losing that quantified estimate.\\

\textit{\textbf{Considering the rate of convergence to equilibrium when $\gamma <1$}}, one may wonder if, at least for initial data close enough to equilibrium, the rate of convergence to equilibrium in the case when $a_i=i^\gamma$ with $\gamma <1$ can be improved to an exponential. Recently, Murray and Pego investigated this rate of convergence only to conclude an algebraic form of decay (see \cite{MurrayPego}). It would be interesting to verify the optimality of this result by determining whether the linearised operator for the equations admits a spectral gap in $\ell^1$ spaces with polynomial weights (in $\ell^1$ spaces with exponential weights, the answer is positive and an estimate of the spectral gap can be found in \cite{JACBL}). The authors believe that no such spectral gap exists for $0 \leq \gamma < 1$, i.e. the algebraic rate of convergence is optimal even for close to equilibrium initial data.\\

\textit{\textbf{Consider the general coagulation and fragmentation equations}} presented in Section \ref{sec:application to general CF}. It seems to the authors that the method used is too crude and is far from optimal. We suspect that the inequality obtained in Theorem \ref{thm:almost cerc for CF} can be improved to deal with the case
$$a_{i,j}=i^\gamma j^\eta+i^\eta j^\gamma$$
and the resulting convergence rate will depend on $\lambda=\gamma+\eta$.\\

The authors are curious to see if some of the aforementioned problems can be addressed with the help of the presented work, and are eager to see possible new functional inequalities arising in connection to the Becker-D\"oring equations, or more general coagulation and fragmentations models.

 \appendix
\section{Additional Computations for the Theory of the Discrete Log-Sobolev With Weights Inequality}\label{app:discrete log-sob} 
We have collected here technical Lemmas from Subsection \ref{sec:discrete log-sob} that we felt would have encumbered the flow of it.
\begin{lem}\label{lemapp:comparison between L and Entropy}
For any sequence $\bm{f}$, we have
\begin{equation*}
\Ent_{\mu}(\bm{f}^2)\leq \mathcal{L}(\bm{f}) \leq \Ent_{\mu}(\bm{f}^2)+2\sum_{i=1}^\infty \mu_i f_i^2. 
\end{equation*}
\end{lem}

\begin{proof}
From the definition of $\mathcal{L}$ the inequality 
$$\Ent_{\mu}(\bm{f}^2)\leq \mathcal{L}(\bm{f}) $$
it trivial. We thus consider the right hand side inequality. For a given sequence $\bm{f}$ and any $\alpha\in \R$ we define
\begin{equation*}\begin{split}
G_{\alpha}(t)&=\sum_{i=1}^\infty \mu_i \pa{tf_i+\alpha}^2\log\pa{\frac{\pa{tf_i + \alpha}^2}{\sum_{i=1}^\infty \mu_i\pa{tf_i+\alpha}^2}} \\
&=2\sum_{i=1}^\infty \mu_i \pa{tf_i+\alpha}^2 \log\abs{tf_i+\alpha} - \pa{\sum_{i=1}^\infty \mu_i \pa{tf_i+\alpha}^2}\log\pa{\sum_{i=1}^\infty \mu_i \pa{tf_i+\alpha}^2},
\end{split}\end{equation*}
and notice that
\begin{equation}\nonumber
G_{0}(t)=t^2\Ent_\mu(\bm{f}^2).
\end{equation}
Next, we define $g(t)=G_{0}(t)+2t^2 \sum_{i=1}^\infty \mu_i f_i^2$ and notice that the inequality we want to prove is equivalent to
\begin{equation}\nonumber
G_{\alpha}(1) \leq g(1).
\end{equation}
for any $\alpha\in \R$. Clearly $G_{\alpha}(t) \leq g(t)$ when $t=0$. Differentiating $G$ we find that
\begin{equation*}\begin{split}
G_{\alpha}^\prime (t) &= 4\sum_{i=1}^\infty \mu_i f_i \abs{tf_i+\alpha} \log\pa{tf_i+\alpha} + 2\sum_{i=1}^\infty \mu_i f_i \pa{tf_i+\alpha}  \\
&\phantom{+++++} -2\pa{\sum_{i=1}^\infty \mu_i f_i \pa{tf_i+\alpha}}\log \pa{\sum_{i=1}^\infty \mu_i \pa{tf_i+\alpha}^2}-2\sum_{i=1}^\infty \mu_i f_i \pa{tf_i+\alpha}
\\
&=4\sum_{i=1}^\infty \mu_i f_i \pa{tf_i+\alpha} \log\abs{tf_i+\alpha} -2\pa{\sum_{i=1}^\infty \mu_i f_i \pa{tf_i+\alpha}}\log \pa{\sum_{i=1}^\infty \mu_i \pa{tf_i+\alpha}^2}
\end{split}\end{equation*}
which satisfies $G_{\alpha}^\prime(0)=0$ for any $\bm{f}$ and $\alpha$, implying that $G_{\alpha}^\prime (0)=g^\prime(0)=0$. As $G$ is defined for any $t\in [0,1]$ we see that it is enough to show that when defined, 
\begin{equation}\nonumber
G_{\alpha}^{\prime\prime}(t) \leq g^{\prime\prime}(t)
\end{equation}
for any $\alpha$. Indeed, 
\begin{equation*}\begin{split}
G_{\alpha}^{\prime\prime}(t)&=4\sum_{i=1}^\infty \mu_i f_i^2\log \abs{tf_i+\alpha}+4\sum_{i=1}^\infty \mu_i f_i^2-2\sum_{i=1}^\infty \mu_i f_i^2\log \pa{\sum_{i=1}^\infty \mu_i \pa{tf_i+\alpha}^2}\\
&\phantom{+++++++ }-4\frac{\pa{\sum_{i=1}^\infty \mu_i f_i \pa{tf_i+\alpha}}^2}{\sum_{i=1}^\infty \mu_i \pa{tf_i+\alpha}^2}\\
&=2\sum_{i=1}^\infty \mu_i f_i^2\log\pa{\frac{\pa{tf_i+\alpha}^2}{\sum_{i=1}^\infty \mu_i \pa{tf_i+\alpha}^2}}+4\sum_{i=1}^\infty \mu_i f_i^2-4\frac{\pa{\sum_{i=1}^\infty \mu_i f_i \pa{tf_i+\alpha}}^2}{\sum_{i=1}^\infty \mu_i \pa{tf_i+\alpha}^2}
\end{split}
\end{equation*}
As 
$$\Ent_\mu(\bm{f}^2) = \sup \br{\sum_{i=1}^\infty\mu_i f_i^2 \log h_i \; ; \; \sum_{i=1}^\infty \mu_i h_i=1}$$
we see that by choosing $h_i=\frac{\pa{tf_i+\alpha}^2}{\sum_{i=1}^\infty \mu_i \pa{tf_i+\alpha}^2} $
$$G_{\alpha}^{\prime\prime}(t) \leq 2 \Ent_{\mu}(\bm{f}^2) + 4\sum_{i=1}^\infty \mu_i f_i^2 = g^{\prime\prime}(t), $$
completing the proof.
\end{proof}

\begin{lem}\label{lemapp: 1 2 and phi}
For all $\bm{f}\in L_{\Phi}$ we have that
\begin{equation}\label{eqapp: 1 2 and phi}
\norm{\bm{f}}_{L^1_\mu}\leq \norm{\bm{f}}_{L^2_\mu} \leq \sqrt{\frac{3}{2}} \norm{\bm{f}}_{L_\Phi}.
\end{equation}
\end{lem}

\begin{proof}
The inequality 
$$\norm{\bm{f}}_{L^1_\mu}\leq \norm{\bm{f}}_{L^2_\mu} $$
is immediate as $\bm{\mu}$ is a probability measure. To show the last inequality we may assume that $\norm{\bm{f}}_{L_\Phi}=1$. Due to Fatou's Lemma we know that if $k_n\underset{n\rightarrow\infty}{\longrightarrow}k >0$ then
$$\sum_{i=1}^\infty \mu_i \Phi\pa{\frac{\abs{f_i}}{k}} \leq \liminf_{n\rightarrow\infty}\sum_{i=1}^\infty \mu_i \Phi\pa{\frac{\abs{f_i}}{k_n}},$$
implying that if $\norm{f}_{L_\Phi}>0$ then
$$\sum_{i=1}^\infty \mu_i \Phi\pa{\frac{\abs{f_i}}{\norm{f}_{L_\Phi}}} \leq 1.$$
In our case, since $\Psi(x)$ is convex we find that
$$1 \geq \sum_{i=1}^\infty \mu_i \Phi(f_i) = \sum_{i=1}^\infty \mu_i \Psi(f_i^2) \geq \Psi\pa{\sum_{i=1}^\infty \mu_i f_i^2}=\Psi\pa{\norm{\bm{f}}^2_{L^2_{\mu}}}. $$
As $\Psi$ is increasing and $\Psi(1.5)  >1$ we conclude that
$$\norm{\bm{f}}^2_{L^2_{\mu}}<\frac{3}{2},$$
yielding the desired result.
\end{proof}

\begin{lem}\label{lemapp: entropy and L_2}
Let $\bm{f}\in L_{\Phi}$. Then
\begin{equation}\label{eqapp: entropy and L_2}
\norm{\bm{f}-\avg{\bm{f}}}_{L^2_\mu}^2 = \frac{1}{2}\lim_{\abs{a}\rightarrow\infty}\Ent_\mu\pa{(\bm{f}+a)^2}
\end{equation}
\end{lem}
\begin{proof}
We start by noticing that
$$\Ent_\mu\pa{(\bm{f}+a)^2}=\sum_{i=1}^\infty \mu_i \pa{f_i^2+2a f_i + a^2}\log\pa{\frac{\pa{1+\frac{f_i}{a}}^2}{\sum_{i=1}^\infty \mu_i \pa{1+\frac{f_i}{a}}^2}}, $$
and continue by assuming that $f_i$ is uniformly bounded, from which the result will follow with an application of an appropriate convergence theorem. There exists $a_0$ such that if $\abs{a}>\abs{a_0}$ we have that $\abs{\frac{f_i}{a}}<\frac{1}{2}$ uniformly in $i$. As on $\left[-\frac{1}{2},\frac{1}{2} \right]$ we have that there exists $C>0$ such that
$$\abs{\log(1+x)-x+\frac{x^2}{2}} \leq C x^3.$$
we conclude that
$$\log\pa{1+2\frac{f_i}{a}+\frac{f_i^2}{a^2}}=\pa{2\frac{f_i}{a}+\frac{f_i^2}{a^2}} - 2\frac{f_i^2}{a^2}+\frac{E_{1,i}}{a^3}=2\frac{f_i}{a}-\frac{f_i^2}{a^2}+\frac{E_{1,i}}{a^3}$$
and 
$$\log\pa{1+2\frac{\avg{\bm{f}}}{a}+\frac{\norm{\bm{f}}_{L^2_\mu}^2}{a^2}}=
2\frac{\avg{\bm{f}}}{a}+\frac{\norm{\bm{f}}_{L^2_\mu}^2}{a^2}-2\frac{\avg{\bm{f}}^2}{a^2}+\frac{E_{2,i}}{a^3},$$
where $E_{1,i},E_{2,i}$ are uniformly bounded in $i$. This implies that
\begin{multline*}\Ent_\mu \pa{(\bm{f}+a)^2}=\sum_{i=1}^\infty \mu_i \pa{f_i^2+2a f_i + a^2}\pa{2\frac{f_i}{a}-2\frac{\avg{\bm{f}}}{a}-\frac{f_i^2}{a^2}-\frac{\norm{\bm{f}}_{L^2_\mu}^2}{a^2}+2\frac{\avg{\bm{f}}^2}{a^2}}\\
+\frac{1}{a}\sum_{i=1}^\infty \mu_i \pa{1+2\frac{f_i}{a}+\frac{f_i^2}{a^2}}\pa{E_{1,i}-E_{2,i}}.\end{multline*}
The last term clearly goes to zero as $\abs{a}$ goes to infinity, so we are only left to deal with the first expression.
\begin{multline*}\sum_{i=1}^\infty \mu_i \pa{f_i^2+2a f_i + a^2}\pa{2\frac{f_i}{a}-2\frac{\avg{\bm{f}}}{a}-\frac{f_i^2}{a^2}-\frac{\norm{\bm{f}}_{L^2_\mu}^2}{a^2}+2\frac{\avg{\bm{f}}^2}{a^2}}=4\norm{\bm{f}}^2_{L^2_\mu}-4\avg{\bm{f}}^2\\
+2a\avg{\bm{f}}-2a\avg{\bm{f}}-\norm{\bm{f}}_{L^2_\mu}^2-\norm{\bm{f}}_{L^2_\mu}^2+2\avg{\bm{f}}^2+\frac{E_{3}}{a}\\
=2\pa{\norm{\bm{f}}_{L^2_\mu}^2-\avg{\bm{f}}^2}+\frac{E_3}{a}.\end{multline*}
This completes the proof as $\norm{\bm{f}-\avg{\bm{f}}}_{L^2_\mu}^2=\norm{\bm{f}}_{L^2_\mu}^2-\avg{\bm{f}}^2$.
\end{proof}

\begin{lem}\label{lemapp: evaluation of norm of c^0 and c^1}
Let $\bm{f}$ be a sequence such that $f_m=0$ for some $m\in\N$. Denote by $\bm{f}^{(0)}=\bm{f} 1\!\!1_{i<m}$ and $\bm{f}^{(1)}=\bm{f} 1\!\!1_{i>m}$. Then
\begin{equation}\label{eqapp: evaluation of norm of c^0 and c^1}
\begin{gathered}
\norm{\avg{\bm{f}^{(0)}}}_{L_\Phi } \leq \abs{\avg{\bm{f}^{(0)}}} \leq \norm{\bm{f}^{(0)}}_{L^2_\mu}\sqrt{\sum_{i=1}^{m-1} \mu_i}\\
\norm{\avg{\bm{f}^{(1)}}}_{L_\Phi } \leq \abs{\avg{\bm{f}^{(1)}}} \leq \norm{\bm{f}^{(1)}}_{L^2_\mu}\sqrt{\sum_{i=m+1}^\infty \mu_i}
\end{gathered}
\end{equation}
\end{lem}
\begin{proof}
We start by noticing that for any constant sequence $\bm{f}=\alpha$ one have
\begin{equation*}\begin{split}\norm{\alpha}_{L_\Phi}&=\inf_{k>0}\br{\sum_{i=1}^\infty \mu_i \Phi\pa{\frac{\abs{\alpha}}{k}}\leq 1}=\inf_{k>0}\br{\Phi\pa{\frac{\abs{\alpha}}{k}}\leq 1}\\
&=\frac{\abs{\alpha}}{\Phi^{-1}(1)} \leq \abs{\alpha},\end{split}\end{equation*}
as long as $\Phi(1) <1$ which is valid in our case. Next we notice that
$$\abs{\avg{\bm{f}^{(0)}}} \leq \sum_{i=1}^{m-1}\mu_i \abs{f_i} \leq \sqrt{\sum_{i=1}^{m-1}\mu_i f_i^2}\sqrt{\sum_{i=1}^{m-1}\mu_i}= \norm{\bm{f}^{(0)}}_{L^2_\mu}\sqrt{\sum_{i=1}^{m-1}\mu_i}.$$
This yields the first inequality and similar arguments yield the second inequality.
\end{proof}

\begin{rem}
As was shown in the proof of Lemma \ref{lemapp: evaluation of norm of c^0 and c^1} one can actually improve the bounds in (\ref{eqapp: evaluation of norm of c^0 and c^1}) by a factor of $\Psi^{-1}(1)$. 
\end{rem}

\begin{lem}\label{lemapp: evaluation of Psi inverse }
For any $t\geq \frac{3}{2}$ one has that
\begin{equation}\label{eqapp: evaluation of Psi inverse }
\frac{1}{3}\frac{t}{\log t} \leq \Psi^{-1}(t) \leq 2 \frac{t}{\log t}.
\end{equation}
\end{lem}
\begin{proof}
We start by noticing that 
$$\Psi\pa{\frac{1}{3}\frac{t}{\log t} } =\frac{1}{3}\frac{t}{\log t}\log \pa{1+ \frac{1}{3}\frac{t}{\log t}}  \leq \frac{1}{3}\frac{t}{\log t}\log \pa{1+ \frac{t}{\log\pa{\frac{27}{8}} }}  $$
$$\leq \frac{1}{3}\frac{t}{\log t}\log \pa{1+ t}. $$
Thus, one notices that if 
$$1+t \leq t^3$$
when $t\geq \frac{3}{2}$, we have that $\Psi\pa{\frac{1}{3}\frac{t}{\log t}} \leq t$, yielding the left hand side of (\ref{eqapp: evaluation of Psi inverse }). This is indeed the case as $g(t)=t^3-t-1$ is increasing on $\left[\frac{1}{\sqrt{3}},\infty \right)$ and $g\pa{\frac{3}{2}}>0$.\\
For the converse we notice that 
$$\Psi\pa{2\frac{t}{\log t}}=2\frac{t}{\log t}\log\pa{1+2\frac{t}{\log t}} \geq t$$
if and only if
$$1+2\frac{t}{\log t} \geq \sqrt{t}.$$
Considering the function $g(x)= \frac{x}{\log x}$ for $x>1$ we see that it obtains a minimum at $x=e$. Thus, for any $x>1$ $g(x) \geq e >1$. We conclude that for $t>\frac{3}{2}$
$$2\frac{t}{\log t} = \sqrt{t}g(\sqrt{t}) \geq \sqrt{t},$$
showing the desired result.
\end{proof}

\section{Additional Useful Computations}\label{app:additional useful}\label{appB}
\begin{lem}\label{lemapp:H(c|Q) minimizes }
For a given coagulation and detailed balance coefficients, $\br{a_i}_{i\in\N},\br{Q_i}_{i\in\N}$, and a given positive sequence $\bm{c}$ with finite mass $\varrho$, we have that for any $z>0$
$$H(\bm{c}|\Q) \leq H(\bm{c}|\Q_z),$$
where $\Q=Q_{\overline{z}}$. 
\end{lem}
\begin{proof}
We have that
$$H(\bm{c}|\Q_z)=\sum_{i=1}^\infty c_i\pa{\log\pa{\frac{c_1}{Q_iz^i}}-1} + \sum_{i=1}^\infty Q_i z^i$$
implying that
$$H(\bm{c}|\Q_{z_1})-H(\bm{c}|\Q_{z_2})=\sum_{i=1}^\infty i c_i \log\pa{\frac{z_2}{z_1}}+\sum_{i=1}^\infty Q_i \pa{z_1^i-z_2^i}.$$
In particular, if $z_2=\overline{z}$ we have that for any $z>0$
$$H(\bm{c}|\Q_z)=H(\bm{c}|\Q)+\varrho \log\pa{\frac{\overline{z}}{z}}+\sum_{i=1}^\infty Q_i \pa{z^i-\overline{z}^i}$$
$$=H(\bm{c}|\Q)+\sum_{i=1}^\infty iQ_i \overline{z}^i \log\pa{\frac{\overline{z}}{z}} +\sum_{i=1}^\infty Q_i z^i\pa{1-\pa{\frac{\overline{z}}{z}^i}}$$
$$=H(\bm{c}|\Q) + \sum_{i=1}^\infty Q_i z^i \pa{\pa{\frac{\overline{z}}{z}}^i \log\pa{\pa{\frac{\overline{z}}{z}}^i }-\pa{\frac{\overline{z}}{z}}^i+1}$$
$$=H(\bm{c}|\Q)+\sum_{i=1}^\infty Q_iz^i \Delta\pa{\frac{\pa{\Q_z}_i}{\Q_i}},$$
where $\Delta(x)=x\log x-x+1>0$ when $x>0$. This completes the proof.
\end{proof}

\begin{lem}\label{lemapp:control for jabin proof}
Let $\br{Q_i}_{i\in\N}$ be a non-negative sequence such that $\lim_{i\rightarrow\infty}\frac{Q_{i+1}}{Q_i}=\frac{1}{r}$ for some $r>0$. Assume that $0<x<r_1<r$. Then
\begin{equation*}
\sum_{i=i_0+1}^\infty iQ_i x^{i-1} \leq C Q_{i_0}x^{i_0},
\end{equation*}
where $C$ is a constant depending only on $\br{Q_i}_{i\in \N}$ and $r_1$. 
\end{lem}
\begin{proof}
Define $\beta_i=\frac{Q_{i+1}}{Q_i}$. We have that $\lim_{i\rightarrow\infty}\beta_i=\frac{1}{r}$, and as such we fan find $l\in\N$ such that for all $i>l$  
$$\Lambda_1 = \sup_{i>l}\beta_i < \frac{1}{r_1}.$$
Denote $\Lambda_2=\sup_{i\leq l}\beta_i$. As for any $i>i_0$ 
$$Q_i=\pa{\prod_{j=i_0}^{i-1}\beta_j}Q_{i_0}$$ 
we see that
$$\sum_{i=i_0+1}^\infty iQ_i x^{i-1}=Q_{i_0}x^{i_0}\sum_{i=i_0+1}^\infty i\pa{\prod_{j=i_0}^{i-1}\beta_j} x^{i-i_0-1} \leq Q_{i_0}x^{i_0}\pa{\Lambda_2\sum_{j=0}^{l-i_0} i\pa{\Lambda_2 r_1}^{j}+\Lambda_1 \sum_{j=l+1-i_0}^\infty i\pa{\Lambda_1 r_1}^{j}}$$
$$\leq Q_{i_0}x^{i_0}\pa{\Lambda_2 \sum_{j=0}^{l} j\pa{\Lambda_2 r_1}^{j}+\Lambda_1 \sum_{j=0}^\infty j\pa{\Lambda_1 r_1}^{j},}$$
completing the proof as $l,\Lambda_1$ and $\Lambda_2$ depend  solely on $\br{Q_i}_{i\in\N}$
\end{proof}

\begin{lem}\label{lemapp:order of B_epsilon_gamma}
Let $\varepsilon>0$ and $\gamma>0$. Denote by
$$B_{\varepsilon,\gamma}=\sum_{i=1}^{\infty}i^\gamma e^{-\varepsilon i}.$$
Then $\varepsilon^{1+\gamma}B_{\varepsilon,\gamma}$ is of order $1$ when $\varepsilon$ goes to zero.
\end{lem}

\begin{proof}
We start by noticing that the function $g_{\varepsilon,\gamma}(x)=x^\gamma e^{-\varepsilon x}$ is increasing in $\left[0,\frac{\gamma}{\varepsilon} \right]$ and decreasing in $\left[ \frac{\gamma}{\varepsilon},\infty \right)$. As such
$$B_{\varepsilon,\gamma} \geq \sum_{i=\left[ \frac{\gamma}{\varepsilon} \right]+1}^\infty i^\gamma e^{-\varepsilon i} \geq \int_{\left[ \frac{\gamma}{\varepsilon} \right]+1}^\infty x^{\gamma}e^{-\varepsilon x}\d x$$
$$=\varepsilon^{-(1+\gamma)}\int_{\varepsilon\pa{\left[ \frac{\gamma}{\varepsilon} \right]+1}}^\infty y^\gamma e^{-y}\d y \geq \varepsilon^{-(1+\gamma)}\int_{\varepsilon}^\infty y^\gamma e^{-y}\d y,$$
showing the lower bound. For the upper bound we notice that
$$B_{\varepsilon,\gamma} \leq \sup_{x\geq 0}g_{\frac{\varepsilon}{e},\gamma}(x) \sum_{i=1}^\infty e^{-\frac{\varepsilon}{2} i} = \pa{\frac{2\gamma}{\varepsilon}}^{\gamma} e^{-\gamma}\frac{e^{-\frac{\varepsilon}{2}}}{1-e^{-\frac{\varepsilon}{2}}}$$
 {which completes the proof since $\sup_{\varepsilon >0}\dfrac{\varepsilon\,e^{-\frac{\varepsilon}{2}}}{1-e^{-\frac{\varepsilon}{2}}} < +\infty.$}\end{proof}

\bibliographystyle{plain}
\bibliography{bibliography}

\end{document}